\documentclass[10pt,prb,preprintnumbers,showpacs,amsmath,amssymb,floatfix,superscriptaddress,twocolumn]{revtex4}
\usepackage{graphicx}
\usepackage{epstopdf}
\usepackage{dcolumn}
\usepackage{bm}
\usepackage{longtable}
\usepackage{graphics}
\usepackage{amssymb}
\usepackage{amsmath}
\usepackage{xspace}
\usepackage{epsfig}
\def \dt{$|\Delta/t|$~}
\def \dtn{$|\Delta/t|$}
\def \Nax{Na$_x$CoO$_2$~}
\def \Naxn{Na$_x$CoO$_2$}
\def \Na{Na$_{0.5}$CoO$_2$~}
\def \Nan{Na$_{0.5}$CoO$_2$}
\def \K{K$_{0.5}$CoO$_2$~}
\def \Kn{K$_{0.5}$CoO$_2$}
\def \Rb{Rb$_{0.5}$CoO$_2$~}
\def \Rbn{Rb$_{0.5}$CoO$_2$}
\def \Ax{$A_x$CoO$_2$~}
\def \Axn{$A_x$CoO$_2$}
\def \A{$A_{0.5}$CoO$_2$~}
\def \An{$A_{0.5}$CoO$_2$}
\newcommand {\ibid}{{\it ibid}. }
\newcommand {\etal}{{\it et al}. }
\newcommand {\etalc}{{\it et al}., }
\newcommand {\etaln}{{\it et al}.}
\begin{document}
\title{Ionic Hubbard model on a triangular lattice for Na$_{0.5}$CoO$_2$, Rb$_{0.5}$CoO$_2$ and
K$_{0.5}$CoO$_2$: Mean-field slave boson theory}
\author{B. J. Powell}
\email{bjpowell@gmail.com} \affiliation{Centre for Organic Photonics and Electronics, School of Mathematics and Physics,
The University of Queensland, Brisbane, Queensland 4072,
Australia}
\author{J. Merino}
\affiliation{Departamento de F\'isica Te\'orica de la Materia
Condensada,
Universidad Aut\'onoma de Madrid, Madrid 28049, Spain}
\author{Ross H. McKenzie}
\affiliation{Centre for Organic Photonics and Electronics, School of Mathematics and Physics,
The University of Queensland, Brisbane, Queensland 4072,
Australia}


\begin{abstract}
We introduce a strongly correlated mean-field theory of the ionic
Hubbard model on the triangular lattice with alternating stripes of
site energy using Barnes-Coleman slave bosons. We study the
paramagnetic phases of this theory at three quarters filling, where
it is a model of \Nan, \Rbn, and \Kn. This theory has two bands
of fermionic quasi-particles: one
of which is  filled or nearly filled and hence weakly correlated;
the other is  half-filled or nearly half-filled and hence strongly
correlated. Further results depend strongly on the sign of the
hopping integral, $t$. The light band is always filled for $t>0$,
but only becomes filled for $|\Delta/t|\geq1.5$ for $t<0$, where
$\Delta$ is the difference in the site energies of the two
sublattices. A metal--charge transfer insulator transition occurs at
$|\Delta/t|=5.0$ for $t>0$ and $|\Delta/t|=8.0$ for $t<0$. In the
charge transfer insulator complete charge disproportionation occurs:
one sublattice is filled and the other in half filled. We compare
our results with exact diagonalisation calculations and experiments
on \Nan, and discuss the relevance of our results to \Rb and \Kn. We propose a resolution of seemingly contradictory
experimental results on \Nan. Many experiments suggest that there is a charge gap,
yet quantum oscillations are observed suggesting the existence of
quasiparticle states at arbitrarily low excitation energies. We
argue that the heavy band is gapped while the light band, which
contains less than 1 charge carrier per 100 unit cells, remains
ungapped.
\end{abstract}

\maketitle

\section{Introduction}

Experiments
on \Axn, where $A$ is Na, Rb, or K, show a wide range of strongly
correlated phases.\cite{OngCavaScience,MPM} 
\Nan, in particular, has attracted much attention following the
discovery of superconductivity when water is intercalated into the
system.\cite{Takada} The phase diagram, with doping $x$, shows many
interesting phases,\cite{OngCavaScience,MPM,Foo} including a
`Curie-Weiss metal', A type antiferromagnetism (ferromagnetic layers
stacked antiferromagnetically) and an `insulating' phase seen only
at $x=0.5$. The latter phase is particularly puzzling as many probes
[including resistivity,\cite{Foo} optical conductivity,\cite{wang}
and angle resolved photoemission spectroscopy (ARPES)\cite{Qian}]
suggest that it is insulating, however, Shubnikov-de Haas
oscillations are also observed,\cite{Balicas} suggesting the state
is metallic. The aims of this paper are to present a simple
variational theory of \An, where $A$=Na, Rb, or K, and to attempt to
reconcile these seemingly contradictory experiments.

In \Ax the Co atoms form a triangular lattice and the simplest model
of the band structure, a single band triangular lattice with nearest
neighbour hopping only, gives good agreement with ARPES experiments
on \Naxn.\cite{ARPES-t-neg} The doping of the system is controlled by
the concentration of $A$ ions, $x$, with the single band being half
filled at $x=0$ and filled at $x=1$. However, the $A$ ions order so
as to minimise the mutual Coulomb
repulsion.\cite{Zandbergen,Roger,Zhang} Therefore, the Coulomb
potential due to the $A$ ions is different at different Co atoms
and, because the $A$ ions are ordered, this gives rise to an ordered
arrangement of potentials at the vertices of the triangular lattice.
Thus an  effective Hamiltonian for \Ax is the ionic Hubbard
model.\cite{MPM}

In addition to the  interest in \Ax the ionic Hubbard model is of
significant interest in its own right. The ionic Hubbard model on
half filled bipartite lattices has attracted interest because it
undergoes a transition from a Mott insulator to a band
insulator.\cite{Egami,Kampf,Garg,Craco,Kancharla,ParisPRL,Bouadim}
Furthermore, away from half filling and on frustrated lattices the
ionic Hubbard model shows a subtle interplay between charge and spin
ordering and metallic and insulating
phases.\cite{Penc,Japaridze,ByczukPRL,Kotliar,JaimeLetter,JaimeLong}
The Hamiltonian of the ionic Hubbard model is
\begin{equation}
\hat {\cal H}=-t\sum_{\langle ij\rangle\sigma} \hat
c^\dagger_{i\sigma} \hat c_{j\sigma} +U \sum_i \hat n_{i \uparrow}
\hat n_{i \downarrow} +\sum_{i\sigma} \epsilon_i \hat n_{i\sigma},
\label{ham1}
\end{equation}
where $t$ is the hopping amplitude between nearest neighbour sites
only, $U$ is the effective on-site Coulomb repulsion between two
electrons,  $\epsilon_i$ is the site energy, $\hat
c^{(\dagger)}_{i\sigma}$ annihilates (creates) an electron on site
$i$ with spin $\sigma$ and $\hat n_{i\sigma}=\hat
c^\dagger_{i\sigma}\hat c_{i\sigma}$. Previous studies of this model
on frustrated lattices included both analytical and numerical
studies of zigzag ladders,\cite{Japaridze,Penc} dynamical mean field
theory (DMFT) studies on infinite dimensional fcc\cite{ByczukPRL}
and two-dimensional triangular lattices\cite{Kotliar} and exact
diagonalisation on small triangular
lattices.\cite{JaimeLetter,JaimeLong} A more extensive discussion of
previous work on the ionic Hubbard model is given in Ref. \onlinecite{JaimeLong}. However, here we
aim to provide a simple variational description that captures as
much of the strongly correlated physics of this model as possible.

The wide range of numerical techniques, described above, that have
been applied to the ionic Hubbard model have not, previously, been
complemented by a commensurate effort to develop simple variational
approaches. The history of the theory of strongly correlated
electrons shows that progress has often been made when accurate
numerical techniques are combined with such variational
calculations. Therefore,  our theory, which can be straightforwardly
generalised to other potential arrangements, lattices, fillings,
etc., is complementary to the previous numerical work.

\begin{figure}
\epsfig{file=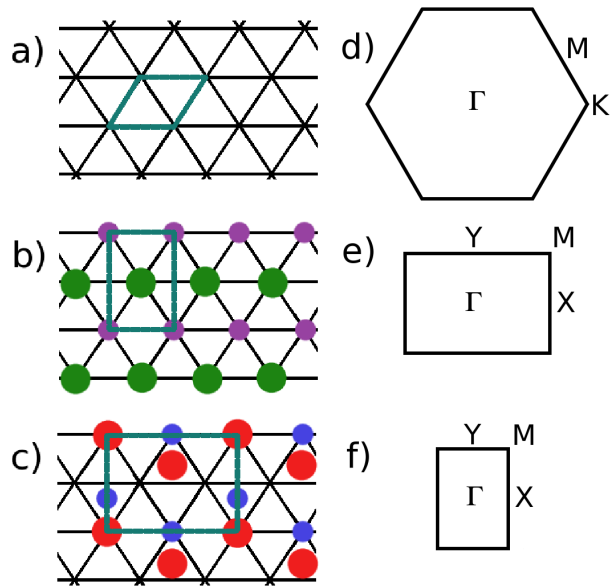,width=8cm} \caption{[Color
online] The various unit cells and Brillouin zones discussed in this paper: (a) the unit
cell of the isotropic triangular lattice; (b) the unit cell of the
ionic Hubbard model [Eq. (\ref{ham})]; and (c) the basal plane  of
the unit cell of \An;  panels (d), (e), and (f) show the Brillouin zones corresponding to the unit cells in panels (a), (b), and (c) respectively. In (b) the different size circles
distinguish the A and B sublattices. In (c) the large red (small
blue) circles indicate the positions of Na ions above (below) the
CoO$_2$ plane. The unit cell of the ionic Hubbard model (b) is twice as
large as that of the Hubbard model of the triangular lattice (a). Further
note that the basal plane of the unit cell of \A is twice as large
as the unit cell of the ionic Hubbard because of the periodicity of the (out
of plane) $A$ ions. Selected high symmetry points are marked in the Brillouin zones. Note that the Brillouin zones are drawn to scale and that the $\Gamma$-point [${\bf k}=(0,0)$] is equivalent in each Brillouin zone.} \label{fig:unit-cells}
\end{figure}

In \Nax many different Na ordering patterns are seen at different values of
$x$,\cite{Zandbergen,Roger,Zhang} and each of these correspond to a different
$\{\epsilon_i\}$.\cite{MPM} In principle, each of these different
$\{\epsilon_i\}$ correspond to a different Hamiltonian. Therefore,
for simplicity and definiteness, we specialise to the Hamiltonian
relevant to $x=0.5$. This model has two sublattices with different
site energies, $\epsilon_i=\Delta/2 $ for A-sites and
$\epsilon_i=-\Delta/2$ for B-sites,\cite{MPM} and with the
sublattices arranged in stripes as shown in Fig.
\ref{fig:unit-cells}b. Thus the Hamiltonian is
\begin{eqnarray}
\hat {\cal H}&=&-t\sum_{\langle ij\alpha\beta\rangle\sigma} \left( \hat c^\dagger_{i\alpha\sigma} \hat c_{j\beta\sigma}  \right)
+U \sum_{i\alpha} \hat n_{i \alpha\uparrow} \hat n_{i\alpha\downarrow}  \notag\\&&
+\sum_{i\sigma} \frac\Delta2 \left(\hat n_{iA\sigma} - \hat n_{iB\sigma}\right), \label{ham}
\end{eqnarray}
where $\hat c^{(\dagger)}_{i\alpha\sigma}$ annihilates (creates) an
electron with spin $\sigma$ in an orbital centred on site $i$
belonging to sublattice $\alpha$ and $\langle ij\alpha\beta\rangle$
indicates that the sum is over nearest neighbours on the appropriate
sublattices only.

Hamiltonian (\ref{ham}) is a natural instance of Hamiltonian
(\ref{ham1}) for initial study on both theoretical and experimental
grounds. Experimentally, the behaviour of, particularly, \Na is very
different from that of \Nax with $x$ either a little larger or
smaller than $0.5$. Further, many of the phenomena observed at $x=0.5$ seem to have a
natural explanation in terms of the ionic Hubbard
model.\cite{JaimeLetter,JaimeLong,MPM} We have previously studied
the quarter-filled ionic Hubbard model with a stripe potential  by
exactly diagonalising small clusters.\cite{JaimeLetter,JaimeLong} We
found a complicated interplay between the charge and spin degrees of
freedom and between insulating and metallic states. We also found
that these calculations provide a possible framework for
understanding a wide variety of experiments on \Nan, \Rbn, and \Kn.
However, the most interesting regime (i.e., $-t\sim\Delta\ll U$, see
section \ref{sect:NaTheory}), in terms of its relevance to
experiments, is also the most challenging to investigate
theoretically. This provides additional motivation to investigate
the simpler mean-field theory presented below. Comparison between
our mean-field theory and these exact diagonalisation calculations
will be made, where possible, below.

The rest of this paper is organised as follows. In the remainder of
this introduction we  briefly review  the most pertinent experiments
on \Na and previous theories aimed at explaining these measurements.
In section \ref{sect:formal} we give the formal details of our
theory and derive a set of self-consistency equations for the ground
state. In section \ref{sect:numeric} we report the results of
numerical
solutions of these equations 
 and compare our results 
with experiments on \A  and previous
theoretical studies of the triangular lattice ionic Hubbard model.
In section \ref{sect:conc} we draw our conclusions. In
appendix \ref{sect:non-int} we describe the band structure of the
model (\ref{ham}) for non-interacting electrons.

\subsection{Experiments on \Nan}

There has been far less experimental work on \K or \Rb than \Nan. Therefore, in this
section, we focus mainly on experiments on \Nan.
The brief review below brings out two particular puzzles that 
theory needs to explain about the ground state. The first is how it
is some properties are consistent with a metallic ground state
and that others are more consistent with an insulating ground state.
The second puzzles concerns how it is that the ground state has a
large magnetic moment but a small amount of charge ordering.

\subsubsection{Charge order on the cobalt ions}
Measurements of the relaxation rates for $^{59}$Co NMR found two
distinct Co sites, consistent  with a charge ordered
state.\cite{ning} However, the authors noted that the degree of
charge disproportionation is rather small. More recent $^{59}$Co and
$^{23}$Na NMR measurements find no signature of differences in the
charge state of the cobalt ions at the two distinct
sites,\cite{bobroff} suggesting that all the Co atoms are in the
charge state
        Co$^{(3.5\pm \eta)+}$, where $\eta < 0.2$
gives an upper bound on the extent of charge ordering.\cite{bobroff}
This is in contrast, to the case of $x=0.7$ or 0.75 for which NMR clearly
detecting a charge disproportionation.\cite{alloul,Julien}
High resolution neutron crystallography detects small differences in
Co-O bond lengths, and an analysis based on bond valence sums, is
consistent with a charge order of $\eta \simeq
0.06$.\cite{williams-argyriou} Very recent NMR measurements were
able to detect a very small charge ordering along the Co(1) chains,
below 51 K,\cite{ning2} but the authors do not discuss the magnitude
of the difference in charge between two sublattices.

\subsubsection{Transport properties}

As the temperature decreases the intralayer resistivity increases
monotonically. No feature is seen at the magnetic ordering
temperature (88 K, see below). Above 51 K, it is weakly temperature
dependent with values of a few m$\Omega$ cm,\cite{Foo} characteristic
of a ``bad'' or incoherent metal which does not have well-defined
quasi-particle excitations.\cite{MPM} Below 51 K, the resistivity
increases significantly, consistent with an activated form with
energy gap of about 10 meV.\cite{Foo} Although, it is usually stated
that the transition at 51 K is a ``metal-insulator'' transition, we
stress that it is really above 51 K \Na is a bad metal  and below 51
K the experimental evidence is not all consistent with the claim
that \Na is an insulator (see section \ref{sect:metal}).

Applying hydrostatic pressure causes the resistivity to decrease and
the  temperature where the resistivity becomes activated  to
increase. Above about 13 GPa the resistivity has a metallic
temperature dependence.\cite{garbarino} In contrast, for $x$=0.75 \Nax
is metallic at ambient pressure but becomes insulating above about
23 GPa.

Resistivity measurements found that the charge gap apparent from the
resistivity was suppressed by magnetic fields parallel to the
layers, larger than about 35 T, with hysteresis between 15 and 40 T,
at low temperatures.\cite{Balicas} A field perpendicular to the
layers does not suppress the ``insulating'' state.\cite{aside}

The thermopower and Hall coefficient have small and weakly
temperature dependent positive values above 100 K.\cite{Foo} However, they
change sign near 88 K, obtain large negative values below 50 K, and
then decrease towards zero at low temperatures. The latter is
distinctly different from the behaviour of conventional
semiconductors and insulators, for which the thermopower and Hall
constant both diverge as the carrier density vanishes with
decreasing temperature. 

\subsubsection{Other evidence for a gap}

Quasi-particle features are only observed in angle resolved
photoemission spectroscopy (ARPES) below about 150 K and reveal a
Fermi surface consistent with a hole-like band for a tight-binding
model on the triangular lattice with an effective hopping integral
of $t_{eff} \simeq -16$ meV,\cite{Qian} about six times smaller than
that predicted by LDA calculations. A gap  begins to open up over
the Fermi surface below about 60 K, and has a magnitude of about
6-11 meV at 10 K.\cite{Qian}

Measurements of the frequency dependence of the conductivity show no
evidence of a Drude peak, consistent with the bad metal behaviour,
seen in the static transport quantities.\cite{wang,MPM} Below about
100 K there is a loss of spectral weight at frequencies below about
 100 cm$^{-1}$,
consistent with the opening of a charge gap. This leads to a peak in
$\sigma(\omega)$ at $\omega \simeq 200$ cm$^{-1}$.\cite{wang} In
comparison for Na$_{x}$CoO$_2$, with $x$ away from 0.5, a Drude peak
is present at low temperatures and the total spectral weight of the
optical conductivity scales with $1-x$.\cite{wang}

\subsubsection{Evidence for a metal}\label{sect:metal}
Shubnikov-de Haas oscillations are observed in the field range of
15-30 T.\cite{Balicas} A fast Fourier transform of the oscillatory
part of the conductivity has peaks at frequencies of 150 T and 40 T.
These frequencies correspond to pockets with cross-sectional areas
of 0.25 \% and 0.06 \% of the area of the undistorted hexagonal
Brillouin zone (cf. Figure 1d).
 The effective mass associated with these orbits are
(in units of the free electron mass) $1.2 \pm 0.1$ and $0.6 \pm
0.5$, respectively. These measurements are extremely surprising
given the range of other evidence suggesting that \Na has an
insulating ground state.


\subsubsection{Magnetic properties}
Muon spin rotation ($\mu$SR) experiments first saw evidence for
magnetic order below about 86 K, and
slight effects at 50 K and 30K.\cite{mendels} Below 88 K, there is a
splitting of NMR lines, consistent with the development of
commensurate antiferromagnetic order.\cite{bobroff} The change in
the resistivity from a bad metal to an activated behaviour at 51 K
has little effect on the magnetic state. Elastic neutron scattering
also detects long-range antiferromagnetic order below 88 K, with a
magnetic moment of $0.26(2)\mu_B$ per magnetic cobalt
ion.\cite{gasparovic,yokoi} The authors interpret this order in
terms of alternating rows of magnetic Co$^{3+}$ and non-magnetic
Co$^{4+}$ ions. (i.e., complete charge order).
 However, we stress that the observation
of such magnetic order does not require charge ordering to be
present. Simple classical arguments  suggest the magnetic moment
should be less than $\mu_B(n_B-n_A)/2$ where $n_\nu$ is the average
number of electrons on the $\nu=A,B$ sublattice.
Hence, the charge disproportionation observed in the crystallography
experiments discussed above, would imply a moment of about
$0.06\mu_B$, about one quarter of the observed value.

Both the bulk magnetic susceptibility and the Knight shift are
weakly temperature dependent, with a peak around 300 K, and a
magnitude of about 5 $\times 10^{-3}$ emu/mol.\cite{bobroff,Foo}
This suggests an antiferromagnetic exchange between magnetic ions of
order a few hundred Kelvin.

\subsubsection{Related materials}
At other values of $x$ in \Naxn, the ordering of
sodium ions has been found to be important for
the existence of various ordered phases.\cite{alloul,alloul2,lang,alloul3,Julien}
No insulating state is seen in the corresponding misfit
cobaltates,\cite{misfit} which supports the claim
 that Na-ordering is necessary
 for the insulating state.
Furthermore, cooling a material at different
rates found that the presence of sodium ordering
can drive an additional magnetic phase transition at $x=0.8$
and 0.85.\cite{schulze}

\subsection{Theories of \Nan}\label{sect:NaTheory}

Choy \etal\cite{Phillips} considered an extended Hubbard model that included the Coulomb interaction between electrons on neighbouring sites, $V$, but neglected the Na-ordering and the `ionic' term in Eq. (\ref{ham}). They argued that $V$ stabilises a charge ordered state and considered an effective low-energy Hamiltonian for this charge ordered state. By fine tuning the parameters in their effective Hamiltonian Choy \etal were able to reproduce the observed temperature dependence of the resistivity and the Hall coefficient.

Lee \etal\cite{PickettChargeSpin,PickettDispro} have studied \Na via
LDA+U calculations that allow for the effects of Na-ordering. They
find a first order metal-charge ordered insulator transition as $U$
is increased. The charge disproportionation and the opening of the
gap both occur at a single phase transition in this weak coupling
theory.

We  have reported exact diagonalisation calculations for Hamiltonian
(\ref{ham}) of finite lattices.\cite{JaimeLetter,JaimeLong} These
studies suggested that, for parameters relevant to \Nan, the system
is a covalent insulator, with little charge disproportionation and
a small gap $\sim|t|$, rather than a charge ordered insulator, with
strong charge disproportionation and a gap $\sim\Delta$. This
prediction is consistent with the experimental measurements of the
charge disproportionation.\cite{bobroff,schulze,williams-argyriou}
Further, these calculations predict a gap of the same order of
magnitude as in seen in the ARPES\cite{Qian} and the
resistivity,\cite{Foo} predict the large moment seen in neutron
scattering,\cite{gasparovic,yokoi} and reproduce the main features
seen in the optical conductivity.\cite{wang}  Zhou and
Wang\cite{ZhouWang} also studied a model that incorporates the
effects of Na-ordering within the Gutzwiller approximation and found
it could explain many of the experimental observations. 

There has been significantly less work on \K and \Rbn. Lee and Pickett\cite{LeeKNa} have reported LDA calculations for \K and compared these with equivalent calculations for \Nan. They found that the $t_{2g}$ band is rather narrower in \K than \Na and that the Fermi surface is more complicated in \K as there are several small pockets along the $X$-$S$ line. They speculated that these pockets, some of which are quite well nested near the $X$ point, may lead to enhanced magnetic tendencies in \Kn.

Clearly an important question is: what parameters in Hamiltonian
(\ref{ham}) correspond to the various possible choices of $A$ in
\An? Various atomistic calculations that address this question have
been presented for $A$=Na, but for other $A$ there are not yet
suitable estimates.
 CAS+DDCI calculations\cite{Landron} on small clusters of \Na
give $-t=0.08-0.14$ eV, $U=2.5-2.8$ eV, and $\Delta=0.16$ eV. For
bulk \Na the LDA yields $\Delta=0.07$ eV \cite{Zhang} and
$t\simeq-0.1$ eV \cite{singh,PickettDispro}, and electrostatic Ewald
calculations give \cite{Roger} $\Delta=0.03-0.05$ eV. Furthermore,
CoO$_2$, which is described by Hamiltonian (\ref{ham}) at half
filling with $\Delta=0$, is observed to be a strongly correlated
metal rather than a Mott insulator.\cite{vaulx} This suggests that
$U \lesssim U_c \sim12-15t$, where $U_c$ is the critical value for
formation of a Mott insulator on the triangular
lattice.\cite{Capone,MPM} Hence, realistic parameters for \Na may be
in the range $10<U/|t|<15$,  and
 $0.5<|\Delta/t|<2$.

The negative sign of $t$ is natural if the hopping between Co sites is
dominated by the contribution from hopping via an intermediate O
site. To leading order this gives
$t=-|t_\textrm{CoO}|^2/(\epsilon_\textrm{Co}-\epsilon_\textrm{O})$,
where $t_\textrm{CoO}$ is the direct hopping between a Co atom and a
neighbouring O atom, $\epsilon_\textrm{Co}$ is the energy of an
electron in a $d$ orbital centred on a Co atom, and
$\epsilon_\textrm{O}$ is the energy of an electron in a $p$ orbital
centred on an O atom (note that one expects that
$\epsilon_\textrm{Co}>\epsilon_\textrm{O}$).

\section{Slave boson theory of strongly-interacting electrons}
\label{sect:formal}

We now investigate paramagnetic phases of the $U\rightarrow\infty$
limit of the ionic Hubbard model by deriving a mean field theory
using Barnes-Coleman  slave bosons.\cite{BC} The $U\rightarrow\infty$ limit great simplifies the analysis and is not unreasonable in spite of our estimate, above, that $10<U/|t|<15$. Our previous exact diagonalization calculations\cite{JaimeLetter,JaimeLong} show that, provided $U\gg |t|$ the physics of this model is insensitive to particular value of $U/|t|$. Specifically, our results for $U=10|t|$ show only minor quantitative differences from those for $U=100|t|$.

\subsection{Theory of the metallic state}

 We begin by making the particle-hole transformation $\hat{c}_{i\alpha\sigma}\rightarrow\hat{h}_{i\alpha{\sigma}}^\dagger$. 
%
We then introduce the slave bosons. Following, for example, Ref.
\onlinecite{LNW}, we write
\begin{eqnarray}
\hat{h}_{i\alpha\sigma}^{\dagger}=\hat{f}_{i\alpha\sigma}^{\dagger}\hat{b}_{i\alpha}+\sum_{\sigma'}\epsilon_{\sigma\sigma'}\hat{f}_{i\alpha\sigma'}\hat{d}_{i\alpha}^{\dagger},
\end{eqnarray}
where $\epsilon_{\sigma\sigma'}$ is the completely antisymmetric
tensor, $\hat{f}_{i\alpha\sigma}^{(\dagger)}$ annihilates (creates)
a fermion with spin $\sigma$ at site $i$ on sublattice $\alpha$,
$\hat{d}_{i\alpha}^{(\dagger)}$ annihilates (creates) a boson,
corresponding to a site that is doubly occupied by holes, at site
$i$ on sublattice $\alpha$, and $\hat{b}_{i\alpha}^{(\dagger)}$
annihilates (creates) a boson, corresponding to a site containing no
holes, at site $i$ on sublattice $\alpha$.
If the number of holes is less than or equal to the number of
lattice  sites and $U=\infty$ there is zero weight for
configurations containing any sites doubly occupied (with holes)  in
the the ground state or any excited states at finite energy above
the ground state. Therefore, $\hat{d}_{i\alpha}$ can be `deleted'
giving us
\begin{eqnarray}
\hat{h}_{i\alpha\sigma}^{\dagger}=\hat{f}_{i\alpha\sigma}^{\dagger}\hat{b}_{i\alpha}.\label{eqn:hole}
\end{eqnarray}
It is clear from Eq. (\ref{eqn:hole}) that the $\hat{f}$'s are
hole-like operators. This transformation is exact, in the limit
$U\rightarrow\infty$, provided we impose the constraint
\begin{eqnarray}
1=\sum_{\sigma}\hat{f}_{i\mu\sigma}^{\dagger}\hat{f}_{i\mu\sigma}+\hat{b}_{i\mu}^{\dagger}\hat{b}_{i\mu} \equiv \hat{Q}_{i\mu} \label{eqn:exact-constraint}
\end{eqnarray}
on the system.



Hence, we find that the Hamiltonian (\ref{ham}) may be written as
\begin{widetext}
\begin{eqnarray}
{\hat{\cal H}}-\mu \hat{N}&=&t\sum_{\langle ij\alpha\beta\rangle\sigma}\hat{f}_{i\alpha\sigma}^{\dagger}\hat{f}_{j\beta\sigma}\hat{b}_{j\beta}^{\dagger}\hat{b}_{i\alpha} 
+\frac\Delta2\sum_{i}\left(\hat{b}_{iA}^{\dagger}\hat{b}_{iA} -\hat{b}_{iB}^{\dagger}\hat{b}_{iB} \right)
- \mu\sum_{i\beta}(1+\hat{b}_{i\beta}^{\dagger}\hat{b}_{i\beta})
\end{eqnarray}

We make a mean-field approximation by replacing the bosons by the
expectation value $q_{ij\alpha\beta}\equiv\langle
\hat{b}_{i\alpha}^{\dagger}\hat{b}_{j\beta} \rangle$, which gives,
\begin{eqnarray}
\hat{\cal H}_\textrm{mf}-\mu \hat{N}&=&t\sum_{\langle ij\alpha\beta\rangle\sigma}\hat{f}_{i\alpha\sigma}^{\dagger}\hat{f}_{j\beta\sigma}q_{ji\beta\alpha}   
+\frac\Delta2\sum_{i}\left(q_{iiAA} -q_{iiBB} \right)
- \mu\sum_{i\beta}(1+q_{ii\beta\beta})
\end{eqnarray}

We now assume that the bosonic mean field is homogeneous
($q_{ij\alpha\beta}=q_{\alpha\beta}$) and introduce the Lagrange
multipliers,  $\lambda_{\alpha}$ to enforce, on average, the local
constraints [via the term
$-\sum_{i\alpha}\lambda_{\alpha}(1-Q_{i\alpha})$]. Thus one finds
that
\begin{eqnarray}
\hat F&\equiv&
\hat{\cal H}_\textrm{mf}-\mu \hat{N} - \sum_{i\alpha}\lambda_{\alpha}(1-\hat{Q}_{i\alpha})  \\\notag
&=&
t\sum_{\langle ij\alpha\beta\rangle\sigma}\hat{f}_{i\alpha\sigma}^{\dagger}\hat{f}_{j\beta\sigma}q_{\beta\alpha}
+\frac\Delta2\sum_{i}\left(q_{AA} -q_{BB} \right)
- \mu\sum_{i\beta}(1+q_{\beta\beta}) 
-\sum_{i\beta}\lambda_{\beta}\left(1-q_{\beta\beta} - \sum_{\sigma}\hat{f}_{i\beta\sigma}^{\dagger}\hat{f}_{i\beta\sigma}\right). \label{eqn:effectiveH}
\end{eqnarray}
Upon performing a Fourier transform one finds that there are two bands, which we denote the bonding (-) and antibonding (+) bands,
\begin{eqnarray}
\hat F &=&
 \label{eqn:early}
\sum_{{\bf k}\sigma}\left(
\begin{array}{ccc}
\hat{f}_{{\bf k}A\sigma} \\
\hat{f}_{{\bf k}B\sigma}
\end{array}
\right)^{\dagger}
\left(
\begin{array}{ccc}
2tq_{AA}\cos k_{x}  + \lambda_{A} & 4tq_{AB}\cos\frac{k_{x}}2\cos\frac{k_{y}}2 \\
4tq_{BA}\cos\frac{k_{x}}2\cos\frac{k_{y}}2 & 2tq_{BB}\cos k_{x}  + \lambda_{B}
\end{array}
\right)\left(
\begin{array}{ccc}
\hat{f}_{{\bf k}A\sigma} \\
\hat{f}_{{\bf k}B\sigma}
\end{array}
\right)
\notag\\ &&
+N\left\{\frac\Delta2\left(q_{AA} -q_{BB} \right) -\sum_{\nu}\left[\mu(1+q_{\nu\nu}) + \lambda_{\nu}(1-q_{\nu\nu}) \right] \right\}
\\&\equiv&
\sum_{{\bf k}\sigma}\left(
\begin{array}{c}
\hat{\psi}_{{\bf k}+\sigma} \\
\hat{\psi}_{{\bf k}-\sigma}
\end{array}
\right)^{\dagger}
\left(
\begin{array}{cc}
E_{{\bf k}+\sigma} & 0 \\
0 & E_{{\bf k}-\sigma}
\end{array}
\right)\left(
\begin{array}{c}
\hat{\psi}_{{\bf k}+\sigma} \\
\hat{\psi}_{{\bf k}-\sigma}
\end{array}
\right)
\notag\\ &&
+
N\left\{\frac\Delta2\left(q_{AA} -q_{BB} \right) -\sum_{\nu}\left[\mu(1+q_{\nu\nu}) + \lambda_{\nu}(1-q_{\nu\nu}) \right] \right\}
\end{eqnarray}
where $k_x$ and $k_y$ are defined in the reduced ($1\times\sqrt3$) Brillouin zone of the model (Fig. \ref{fig:unit-cells}e), $N=\sum_{i}$ is the number of unit cells, i.e., half the number of lattice sites, and the dispersion relations, $E_{{\bf k}\pm\sigma}$, of the `quasiholes', which are destroyed (created) by $\hat{\psi}_{{\bf k}\pm\sigma}^{(\dagger)}$, are given by
\begin{eqnarray}
E_{{\bf k}\pm\sigma} &=&
2tq_{+}\cos k_{x}+\lambda_{+}
\pm\sqrt{\left[2tq_{-}\cos k_{x}  + \lambda_{-} \right]^{2}+16t^{2}|q_{AB}|^{2}\cos^{2}\frac{k_{x}}2\cos^{2}\frac{k_{y}}2}
\end{eqnarray}
where $q_{+} = \frac12(q_{AA}+q_{BB})$,
$q_{-} = \frac12(q_{AA}-q_{BB})$,
$\lambda_{+} = \frac12(\lambda_{A}+\lambda_{B})$, and
$\lambda_{-} = \frac12(\lambda_{A}-\lambda_{B})$.

Hence we find that
\begin{eqnarray}
\hat F
&=&
\sum_{{\bf k}\alpha\sigma}
\left\{ 2tq_{+}\cos k_{x}+\lambda_{+} +\alpha\sqrt{\left[2tq_{-}\cos k_{x}  + \lambda_{-} \right]^{2}+16t^{2}|q_{AB}|^{2}\cos^{2}\frac{k_{x}}2\cos^{2}\frac{k_{y}}2} \right\}\hat{\psi}_{{\bf k}\alpha\sigma}^{\dagger}\hat{\psi}_{{\bf k}\alpha\sigma}  \notag\\ &&
+2N\left[\left(\frac\Delta2+\lambda_-\right)q_{-}  - \mu(1+q_{+})-\lambda_+(1-q_{+})  \right],
\end{eqnarray}
where $\alpha\in\{\pm1\}$.

It is interesting to compare this with the solution of the model with $U=0$ [see Appendix \ref{sect:non-int} and particularly Eq. (\ref{eqn:dispersion})].
This enables a straightforward identification of the physical meaning of the mean fields $q_{\pm}$ and $q_{AB}$ and the Lagrange multipliers $\lambda_{\pm}$, see table \ref{table:meanings}.

\begin{table*}
\caption{Physical meaning of the mean fields, $q_{i}$, and the Lagrange multipliers, $\lambda_{\pm}$. Note that one can also interpret the quasiparticle weights as the inverse of the corresponding  effective mass.}
\begin{center}
\begin{tabular}{lcc} \hline\hline
Parameter & Alternative symbol & meaning \\ \hline
$q_{+}$ & $Z_{+}$ & uniform intra-chain quasiparticle weight \\
$q_{-}$ & $Z_{-}$ & anisotropic intra-chain quasiparticle weight \\
$|q_{AB}|$ & $Z_{AB}$ & inter-chain quasiparticle weight \\
$q_{AA}$ & $Z_{AA}$ & A-sublattice intra-chain quasiparticle weight \\
$q_{BB}$ & $Z_{BB}$ & B-sublattice intra-chain quasiparticle weight \\
$-\lambda_{+}$ & $\mu^{*}$ & effective quasihole chemical potential \\
$-\lambda_{-}$ & ${\Delta^*}/2$ & effective ionic potential\\
\hline\hline
\end{tabular}
\end{center}
\label{table:meanings}
\end{table*}%

It follows from the the approximation that the bosonic mean-field is homogeneous and the constraint of one particle per site that, at three-quarters filling, we need only
consider a two site model (i.e., one unit cell), which must contain
exactly one boson. Thus, the bosonic wavefunction may be written as
\begin{equation}
|\Psi\rangle = \prod_{i}\left( \upsilon e^{i\theta} \hat{b}_{iA}^{\dagger} + \sqrt{1-\upsilon^{2}} \hat{b}_{iB}^{\dagger} \right) |0\rangle,
\end{equation}
where $\upsilon$ is a real number between 0 and 1. 
Thus
\begin{subequations}
\begin{eqnarray}
q_{AA}&=&\langle \Psi| \hat{b}_{iA}^{\dagger} \hat{b}_{iA} |\Psi\rangle = \upsilon ^{2}\\
q_{BB}&=&\langle \Psi| \hat{b}_{iB}^{\dagger} \hat{b}_{iB} |\Psi\rangle = 1 - \upsilon ^{2}\\
q_{AB}&=&\langle \Psi| \hat{b}_{iA}^{\dagger} \hat{b}_{iB} |\Psi\rangle = \upsilon (1-\upsilon) e^{-i\theta}
=\sqrt{q_{AA}q_{BB}}e^{-i\theta} = \sqrt{q_{+}^{2}-q_{-}^{2}}e^{-i\theta}.
\end{eqnarray}
\end{subequations}
Therefore, the total energy of the model is
\begin{eqnarray}
F\equiv\left\langle\hat F\right\rangle
&=&\sum_{{\bf k}\alpha\sigma}
\left\{ 2tq_{+}\cos k_{x}+\lambda_{+} +\alpha\sqrt{\left[2tq_{-}\cos k_{x}  + \lambda_{-} \right]^{2}+16t^{2}(q_{+}^{2}-q_{-}^{2})\cos^{2}\frac{k_{x}}2\cos^{2}\frac{k_{y}}2} \right\}n_{{\bf k}\alpha\sigma} \notag
\\ &&+ 2N\left[\left(\frac\Delta2+\lambda_-\right)q_{-}  - \mu(1+q_{+})-\lambda_+(1-q_{+})  \right],
\label{eqn:free-energy}
\end{eqnarray}
where $n_{{\bf k}\alpha\sigma}\equiv \langle \hat{\psi}_{{\bf
k}\alpha\sigma}^{\dagger}\hat{\psi}_{{\bf k}\alpha\sigma}\rangle$.
The phase  $\theta$ does not change the energy, thus, the
solution has a U(1) degeneracy above that found in the slave boson
mean field theory of the Hubbard model\cite{LNW} due to the two
sublattice structure of the problem.

It is straightforward to show that $q_{+}=x=1/2$ follows from the requirement
 that $-{\partial F}/{\partial \mu} =N_{e}$,
where the total number of electrons is
$N_{e}=\sum_{i\mu\sigma}\hat{c}_{i\mu\sigma}^{\dagger}\hat{c}_{i\mu\sigma}=2N(1+x)$
for $A_{x}$CoO$_{2}$ and we have specialised to $x=1/2$, which is
the relevant filling. This result fits with our intuition that, on average, there
should be one boson (i.e., one doubly occupied site) per unit cell.

Similarly, $(2N)^{-1}\sum_{{\bf k}\alpha\sigma}
n_{{\bf k}\alpha\sigma}=(1-q_{+})=\frac12$ follows from the requirement that $F$ is an extremum with respect to $\lambda_+$.
Again, this fits with our intuition that there is, on average, one quasihole
fermion (singly occupied site) per unit cell. 

Three more complicated intertwined self-consistency conditions can also be derived:
\begin{subequations} \label{eqn:selfconst}
\begin{eqnarray}
\lambda_{-}
=-\frac\Delta2-\frac1{2N} \sum_{{\bf k}\alpha\sigma}
\alpha\left\{\frac{\left[2tq_-\cos k_x  + \lambda_- \right]2t\cos k_x - 16t^2q_{-}\cos^{2}\frac{k_x}{2}\cos^{2}\frac{k_y}{2}}{\sqrt{\left[2tq_-\cos k_x  + \lambda_- \right]^2+16t^2(q_{+}^{2}-q_{-}^{2})\cos^{2}\frac{k_x}{2}\cos^{2}\frac{k_y}{2}}} \right\}n_{{\bf k}\alpha\sigma} \label{eqn:lambda_--self-const}
\end{eqnarray}
follows from the requirement that $F$ is a minimum with respect to $q_-$;
\begin{eqnarray}
q_{-}
= -\frac1{2N} \sum_{{\bf k}\alpha\sigma}
\alpha\left\{\frac{2tq_-\cos k_x  + \lambda_-}{\sqrt{\left[2tq_-\cos k_x  + \lambda_- \right]^2+16t^2(q_{+}^{2}-q_{-}^{2})\cos^{2}\frac{k_x}{2}\cos^{2}\frac{k_y}{2}}} \right\}n_{{\bf k}\alpha\sigma} \label{eqn:q_--self-const}
\end{eqnarray}
follows from the requirement that $F$ is an extremum with respect to $\lambda_-$; and
\begin{eqnarray}
\mu
=\sum_{{\bf k}\alpha\sigma}
\left\{ 2t\cos k_{x} +\alpha\frac{16t^2q_{+}\cos^{2}\frac{k_x}{2}\cos^{2}\frac{k_y}{2}}{\sqrt{\left[2tq_-\cos k_x  + \lambda_- \right]^2+16t^2(q_{+}^{2}-q_{-}^{2})\cos^{2}\frac{k_x}{2}\cos^{2}\frac{k_y}{2}}} \right\}n_{{\bf k}\alpha\sigma}
 +\lambda_+ \notag\\
\end{eqnarray}
\end{subequations}
follows from the requirement that $F$ is a minimum with respect to $q_+$.

\end{widetext}

In order to compare our results with ARPES experiments on
Na$_{0.5}$CoO$_2$ and K$_{0.5}$CoO$_2$ we calculate the spectral
function, $A({\bf k},\omega)\equiv\frac1\pi{\it
ImTr}G_{\mu\nu\sigma\sigma'}({\bf k},\omega)$, where
$G_{\mu\nu\sigma\sigma'}({\bf k},\omega)$ is the electronic
propagator. In real space and time the one electron propagator is
given by
\begin{eqnarray}
iG_{ij\mu\nu\sigma\sigma'}(t)&\equiv&\langle\hat
c_{i\mu\sigma}^\dagger(t)
c_{j\nu\sigma'}(0)\rangle\notag\\&=&q_{\mu\nu}\langle\hat
f_{i\mu\sigma}(t) f_{i\mu\sigma'}^\dagger(0)\rangle
\end{eqnarray}
Fourier transforming and taking the trace, one finds that
\begin{eqnarray}
A({\bf k},\omega)=2\sum_\alpha Z_{{\bf k}\alpha}[1-n_F(E_{{\bf
k}\alpha\sigma})]\delta(E_{{\bf
k}\alpha\sigma}-\omega),\label{eqn:ARPES}
\end{eqnarray}
where $n_F$ is the Fermi function,
\begin{eqnarray}
Z_{{\bf
k}\alpha}=q_++\alpha\cos2\theta_{\bf k}
\end{eqnarray}
is the quasiparticle weight, and
\begin{eqnarray}
\tan\theta_{\bf k}= \frac{E_{{\bf k}+\uparrow}-2t(q_++q_-)\cos
k_x+\lambda_++\lambda_-}{4t\sqrt{q_+^2-q_-^2}\cos\frac{k_x}2\cos\frac{k_y}2}.
\end{eqnarray}
Therefore, we find that, in contrast to mean field slave boson
theories of other models, the quasiparticle weight is momentum
dependent in the theory of the ionic Hubbard model. However, this
does not mean that all non-local correlations induced by the on-site
Coulomb repulsion  are included in the present approach.

\subsection{Theory of the insulating state}

With any minimisation over a finite parameter space it is important
to check explicitly whether any of the end points are the global
minima.\cite{Boas} Therefore we must consider the special case
$q_-=\pm1/2$ separately from the general set of self consistency
conditions [Eqs. (\ref{eqn:selfconst})].
%
Let us consider the case of large (how large will be determined later) positive $\Delta$ such that $q_-=-1/2$, recall that $q_+=1/2$. Eq. (\ref{eqn:early}) then yields
\begin{eqnarray}
F&=&
\sum_{{\bf k}\sigma}\big[(2t\cos k_x +\lambda_B)n_{{\bf k}B\sigma}
+\lambda_A n_{{\bf k}A\sigma}\big] \notag \\&&
-N\left(\frac\Delta2+3\mu+\lambda_A\right).
\end{eqnarray}
Therefore $-{\partial F}/{\partial\mu}=3N$, as is required to ensure
three electrons per unit cell. Further, one finds that $n_{{\bf
k}B\sigma}=0$ for all ${\bf k}$ and $\sigma$ from the requirement
that $F$ is an extremum with respect to $\lambda_B$,
which we expect from $q_B=1$ and the constraint of one particle per cite. Finally,
$\sum_{{\bf k}\sigma}n_{{\bf k}A\sigma}=N$ due to the requirement
that $F$ is an extremum with respect to $\lambda_A$, which is simply
the expected constraint that the A sublattice is half filled with
quasiholes. Note, however, that because of the infinite $U$ or,
equivalently, the constraint of one particle per site, the half
filled A sublattice is insulating. This is also reflected in the
fact that $q_{AA}=q_{AB}=0$, i.e. there is zero quasiparticle weight
for fermions on the A sublattice. So the state $q_A=0$, $q_B=1$ (or,
equivalently, $q_+=1/2$, $q_-=-1/2$) describes a charge transfer
insulator, which has the characteristics of  both Mott (A
sublattice) and band (B sublattice) insulators. Thus, we find that,
\begin{eqnarray}
\frac{F}{N}&=&-\frac\Delta2-3\mu,
\end{eqnarray}
which is clearly just the classical energy of paramagnetic insulating state.

\subsection{Comparison with the empirical theory of Choy \etaln}\label{sect:Choy}

As discussed in section \ref{sect:NaTheory}, Choy \etaln\cite{Phillips}  have studied an empirical model Hamiltonian for holes and doublons motivated by the presumed low energy processes about an antiferromagnetically ordered charge transfer insulating state. This empirical model gives good agreement with the measured temperature dependence of both the resistivity and the Hall coefficient. It is therefore interesting to compare our Eq. (\ref{eqn:effectiveH}) with the Hamiltonian studied by Choy \etal [their Eq. (2)].

Upon making the following notational changes to and substitutions
into Eq. (\ref{eqn:effectiveH}): $tq_{AA}\rightarrow -t_{ij}^d$;
$tq_{BB}\rightarrow t_{ij}^h$; $\lambda_-\rightarrow V+\tilde{J}$;
$tq_{AB}\rightarrow t$;  $\hat f_{iB\sigma}\rightarrow h_{i\sigma}$;
$\sum_\sigma\hat f_{iA\sigma}^\dagger\hat
f_{jA\sigma}=\delta_{ij}-\hat b_{iA}^\dagger\hat b_{jA}$; $\hat
b_{iA}\rightarrow d_i$; $\hat b_{iA}^\dagger\hat f_{jB\sigma}=\hat
f_{iA\sigma}\hat f_{jB\sigma}$ and neglecting constant terms one
finds that the two Hamiltonians are identical. Notice, in
particular, that the renormalised ionic potential, $\lambda_-$, in
our theory plays the role of the inter-site interactions,
$V+\tilde{J}$, in that of Choy \etaln.  Also their non-magnetic
regime ($\alpha=0$) corresponds to $q_-=0$ in our slave boson
theory.  The ground state studied by Choy \etal corresponds to the
charge transfer insulator we only find for $|\Delta/t|\gg1$. Thus,
at least in this regime, the two theories appear to be equivalent.
In order to get agreement with the experimentally measured
temperature dependence of the resistivity and the Hall coefficient
Choy \etal require a specific temperature dependence of (in our
notation) $q_{AA}$, $q_{BB}$, and $\lambda_-$. Although we will not
present finite temperature calculations below, such a temperature
dependence is a quite reasonable expectation for our mean field
theory. Indeed slave boson mean-field theories of other strongly
correlated models have boson mean-fields, similar to $q_-(T)$, that
decrease monotonically with increasing temperature and vanish at
some coherence temperature, $T^*\ll T_F^0$, the Fermi temperature of
the non-interacting system.\cite{Spalek,Burdin}



\section{Numerical results}\label{sect:numeric}

\begin{figure}
\epsfig{file=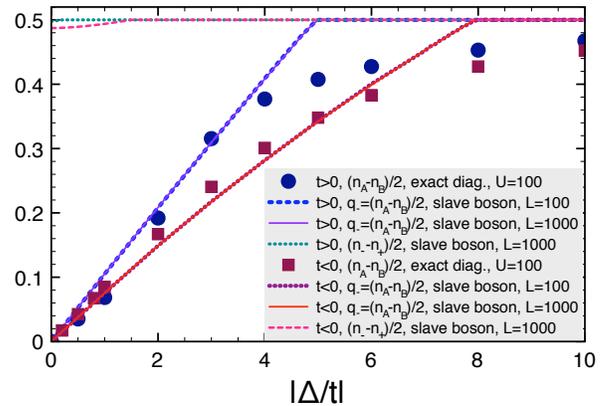,width=8cm} \caption{[Color
online] Charge disproportionation as a function of \dt for both
signs of $t$. Here we compare the real-space charge
disproportionation, $n_A-n_B$, calculated by exact
diagonalisation\cite{JaimeLetter} with the results from the slave
boson calculations where $q_-=(n_A-n_B)/2$. The two methods are in
excellent quantitative agreement for small \dtn, but at large \dt
the slave boson theory predicts complete charge disproportionation,
i.e., one sublattice becomes completely filled. In the exact
diagonalisation results complete charge disproportionation is only
approached asymptotically as $|\Delta/t|\rightarrow\infty$. In the
slave boson theory the system is metallic except when there is
complete charge disproportionation. In the exact diagonalisation
calculations a gap opens when there is only a small charge transfer
between the two sublattices.  We also plot the reciprocal space
(band) charge disproportionation, $n_--n_+$. We see that the MIT
does not occur at the same \dt as the bonding band becomes filled,
in spite of this implying that the antibonding band is half filled
and the fact that $U=\infty$. This emphasises the fact that the
half filled band is \emph{not} equivalent to a half filled 1D chain.
We plot $q_-$ (similar results are found for other parameters) for
two different sized ($L\times L$) reciprocal space meshes. In the
$L=1000$ case the convergence criteria is also tightened by a factor of
50, which demonstrates that the calculations are well converged.
}
\label{fig:q_-}
\end{figure}

We have solved the self consistent equations (\ref{eqn:selfconst}) on an $L\times L$
reciprocal space mesh for $L=100$ and 1000. For the $L=1000$ solutions we
also tightened the convergence criteria by a factor of 50. We find
that the solutions are well converged as the difference between the
$L=100$ and $L=1000$ self consistent solutions is no larger than a
few parts in a thousand for any of the parameters. By way of an example, in Fig.
\ref{fig:q_-} we report both sets of results, it can be seen the
difference in the results is significantly smaller than the width of
the lines. Therefore, we report only the $L=1000$ results in the
subsequent figures.

\subsection{Charge disproportionation and the metal-charge ordered insulator transition}

In Fig. \ref{fig:q_-} we plot the real space charge
disproportionation, $q_-=(n_A-n_B)/2$, where
$n_\mu=\langle\sum_{i\sigma}\hat c_{i\mu\sigma}^\dagger \hat
c_{i\mu\sigma}\rangle$ is the average number of \emph{electrons} per
sublattice site, against \dtn. It can be seen that the results are
strongly dependent on the sign of $t$. The system is
metallic for $|q_-|<1/2$; $|q_-|=1/2$ corresponds to exactly one
fermion on each site of the A-sublattice and exactly one boson on
each site of the B-sublattice. The metal-insulator transition (MIT)
occurs at $|\Delta/t|=8.0$ for $t<0$ and $|\Delta/t|=5.0$ for $t>0$.

We also plot the reciprocal space charge disproportionation (band
filling) in Fig. \ref{fig:q_-}. For $t>0$ the bonding band is filled
for all non-zero $\Delta/t$ (for $\Delta=0$  the unit cell is halved
so there is only a single band). For $t<0$ the bonding band is only
partially filled for $|\Delta/t|<1.5$. As we will discuss in detail
below, the quasiparticles in the bonding (-) band are significantly
lighter than those in the, nearly half-filled, antibonding (+) band.
It is interesting to observe therefore that for $t<0$,
$1.5<|\Delta/t|<8.0$ and $t>0$, $0<|\Delta/t|<5.0$ the antibonding
band is half-filled and, even though $U=\infty$, the system remains
metallic. At first sight this is rather surprising as the system is
two-dimensional and even quasi-one-dimensional. However, because the
bands arise from the hybridisation of the A and the B sublattices
the usual real space arguments that explain the Mott insulating
state do not apply here. There is no interaction between the
quasiparticles other than the real space constraint of one fermion
(or boson) per site. Therefore, unless there is complete charge
disproportionation the system remains metallic.

It is interesting to contrast the results of the slave boson
calculations with our previous exact diagonalisation results for
finite clusters with $U=100|t|$.\cite{JaimeLetter,JaimeLong} At
least at low \dtn, the slave boson theory gives excellent agreement
with the exact diagonalisation for some physical quantities, e.g.,
the level of charge disproportionation (cf. Fig. \ref{fig:q_-}).
But, there is an important qualitative difference between the slave
boson calculations and the exact diagonalisation results: the nature
of the insulating state. In the slave boson calculations an
insulating state is only realised when there is complete charge
disproportionation leaving one sublattice completely filled and the
other half-filled. The filled sublattice acts as a band insulator
and, because $U=\infty$, the half filled sublattice becomes a Mott
insulator. This charge transfer insulator is very different from the
covalent insulating state predicted from the exact diagonalisation
calculations.\cite{JaimeLetter,JaimeLong} These calculations predict
that the insulating state occurs at quite small values of \dt in
spite of there being only rather weak charge disproportionation.
This insulating state depends crucially on the
hybridisation between the two sublattices and is analogous to a
covalent insulator.\cite{Sarma} Our mean field theory neglects
non-local correlations that are included in the exact diagonalisation
calculations and cause the covalent insulating state. Preliminary results suggest that DMFT, which also
neglects non-local correlations but is more sophisticated than our
mean field theory, cannot describe the covalent insulator phase either.

\begin{figure}
\epsfig{file=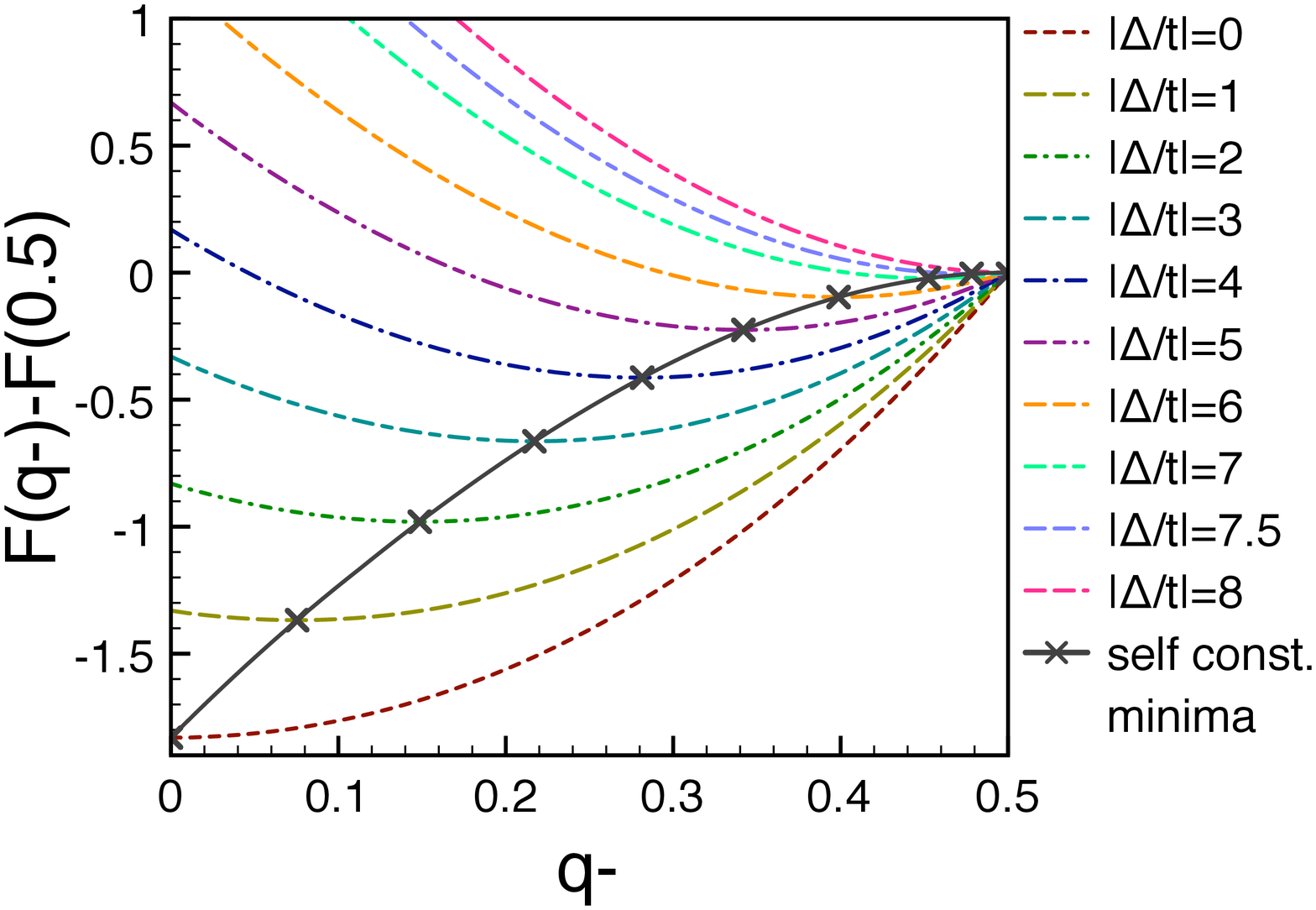,width=8cm} \caption{[Color
online] The ground state energy, $F$, relative to that in the
insulating state for the same value of $\Delta$ as a function of
$q_-$ for various values of $\Delta$ and $t<0$. The solid black line
indicates the free energy found at the self consistent solution,
with crosses marking the data for values of $\Delta$ at which we
plot the full $F(q_-)$ curve. This plot indicates that the self
consistent solutions are indeed minima of $F$ with respect to the
variation of $q_-$ [note that Eq. (\ref{eqn:lambda_--self-const})
only requires that the self consistent solution is a turning point
in $F$ with respect to $q_-$, whereas, physically, we seek the
minimum]. These results also strongly suggest that the metal
insulator transition in the slave boson theory is second order as no
double well structure is seen in $F(q_-)$.} \label{fig:F-t<0}
\end{figure}

\begin{figure}
\epsfig{file=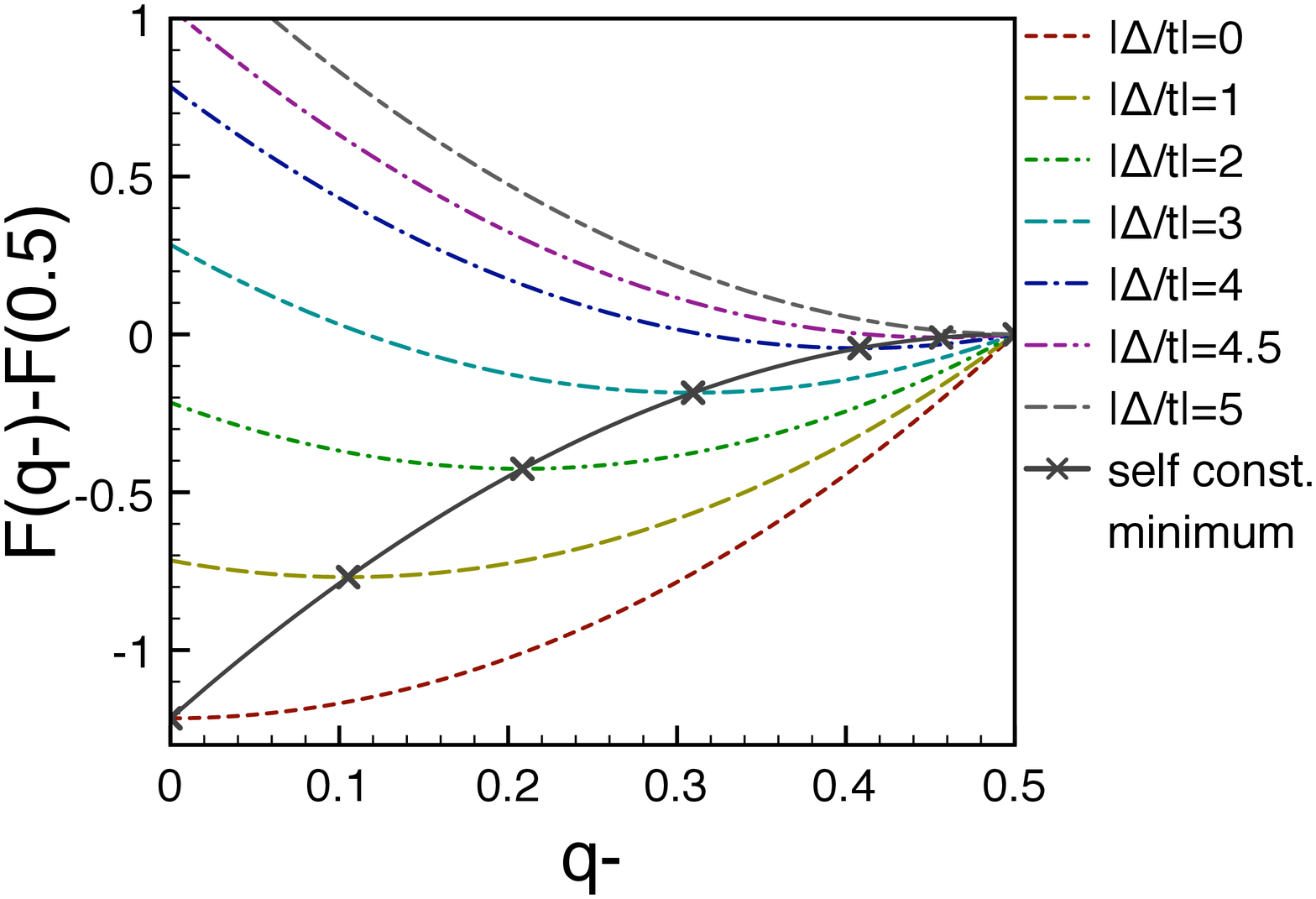,width=8cm} \caption{[Color
online] Same as Fig. \ref{fig:F-t<0} but for $t>0$. Again this shows
that the self consistent solutions are true minima, and that the
metal-charge ordered insulator transition is second order.}
\label{fig:F-t>0}
\end{figure}

In Figs. \ref{fig:F-t<0} and \ref{fig:F-t>0} we plot the ground
state energy, $F$, as a function of $q_-$ for various values of \dt
and the self consistently calculated curve of $F$ against $q_-$ with
\dt as a parametric parameter. It can be seen from these plots that
the self consistent solutions are the true minima. Further, the
absence of double well structures in the plots of $F$ against $q_-$
strongly suggest that the MIT is a second order phase transition in
the slave boson theory.

\subsection{Band structure of the fermionic quasiholes}\label{sect:bs-quasiholes}

We now turn to analyse the properties of the fermionic
quasiholes. As these are non-interacting particles one can
straightforwardly calculate the properties of the quasiholes. It is interesting to compare these results with the band structure of the non-interacting ($U=0$) ionic Hubbard model, which we discuss in Appendix \ref{sect:non-int}.

\begin{figure}
\epsfig{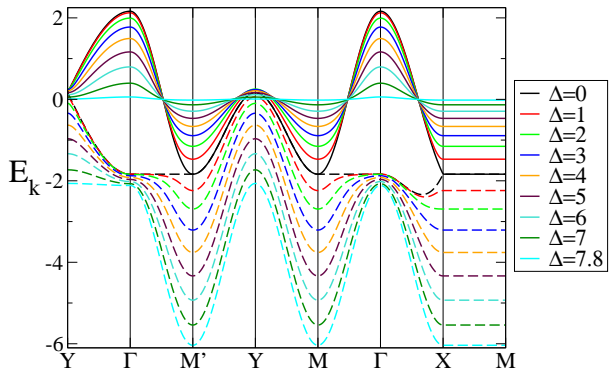} \caption{[Color online]
Band structure of the fermions for $t<0$. The band structure is not unlike that of the
non-interacting system (Fig. \ref{bs-lt}). However, for the
fermions, increasing \dt has a much stronger effect than it does for
the non-interacting electrons. In particular as \dt is increased the
antibonding band (solid lines) is flattened and eventually becomes
flat when the metal-insulator transition is reached at
$|\Delta/t|=8.0$. The high symmetry points of the first Brillouin
zone are shown in Fig. \ref{fig:unit-cells}e. Also see the online movie.
\cite{mov}} \label{fig:bs-t<0}
\end{figure}

Before discussing the band structure in detail we should point out that the bands are, to some extent, an artefact of the mean field approximation. In the exact solution correlations that are not captured by the mean field approximation may destroy the description of the system in terms of extended bands (cf. Ref. \onlinecite{JaimeLong}). Nevertheless, if the mean field theory does provide a reasonable description of the ionic Hubbard model then one expects that some of the features of the band structure described below do survive in the exact solution. Further, the bands are well defined in the mean-field slave-boson theory. Therefore, it is legitimate to calculate their properties in this context and this is extremely useful for building intuition about this problem.  

\begin{figure}
\epsfig{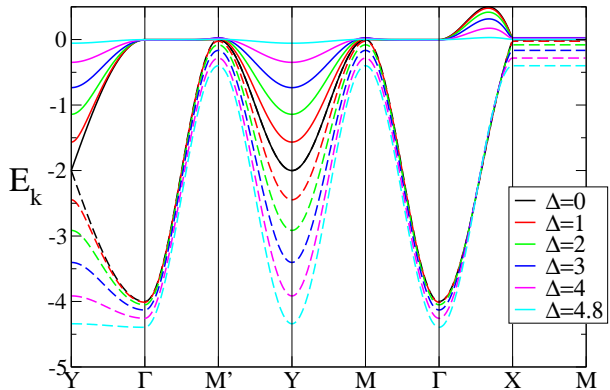} \caption{[Color online]
Band structure of the fermions for $t>0$. There are significant differences from the band
structure of $t<0$ (Fig. \ref{fig:bs-t<0}). In particular the
metal-insulator transition occurs at only $|\Delta/t|=5.0$.  Also see the online movie.
\cite{mov}}
\label{fig:bs-t>0}
\end{figure}

Figs. \ref{fig:bs-t<0} and  \ref{fig:bs-t>0} show how the
dispersion relations of the fermionic quasiholes vary with \dt for each sign of $t$ (also see the
animations online\cite{mov}). These should be
compared with the equivalent plots (Figs. \ref{bs-lt} and
\ref{bs-gt}) for the non-interacting system. It can be seen that at
$\Delta=0$, for both signs of $t$, the quasiparticle bands are
significantly narrower than the equivalent non-interacting electron
bands. As \dt increases the antibonding band narrows and eventually
becomes flat at the MIT. While the bonding band broadens somewhat as \dt
increases.

\begin{figure*}
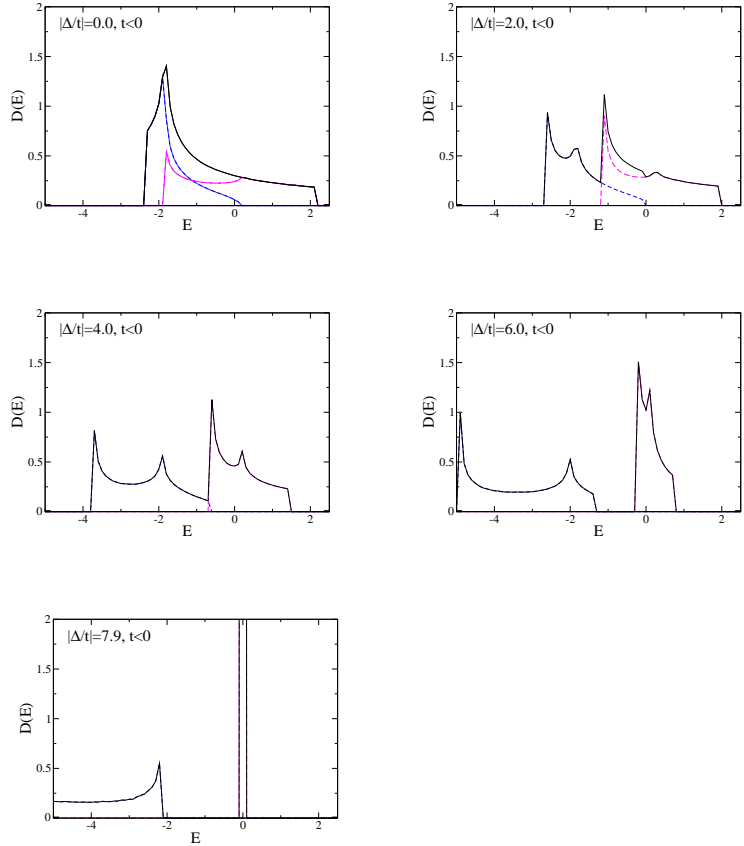

\epsfig{file=DOS_tlt0_Delta_eq0p000.eps,width=4.25cm} \hspace{1cm}
\epsfig{file=DOS_tlt0_Delta_eq2p000.eps,width=4.25cm}\vspace*{1cm}\\
\epsfig{file=DOS_tlt0_Delta_eq4p000.eps,width=4.25cm} \hspace{1cm}
\epsfig{file=DOS_tlt0_Delta_eq6p000.eps,width=4.25cm}\vspace*{1cm}\\
\epsfig{file=DOS_tlt0_Delta_eq7p900.eps,width=4.25cm}\hspace{5.25cm}
\caption{[Color online] Density of states (in units
where $|t|=a=1$) of the fermions for $t<0$.
 Dashed (blue and pink) lines indicate the
contributions of the individual bands. The narrowing of the antibonding band, as $|\Delta/t|$ increases, is clearly visible from these plots, while it can be seen that the bonding band actually becomes a little wider as \dt increases.
Clearly, the DOS is very different from the non-interacting case, cf. Fig. \ref{dos-lt}, and these difference become greater as \dt increases. Also see the online movie.\cite{mov}} \label{fig:DOS-t<0}
\end{figure*}

\begin{figure*}
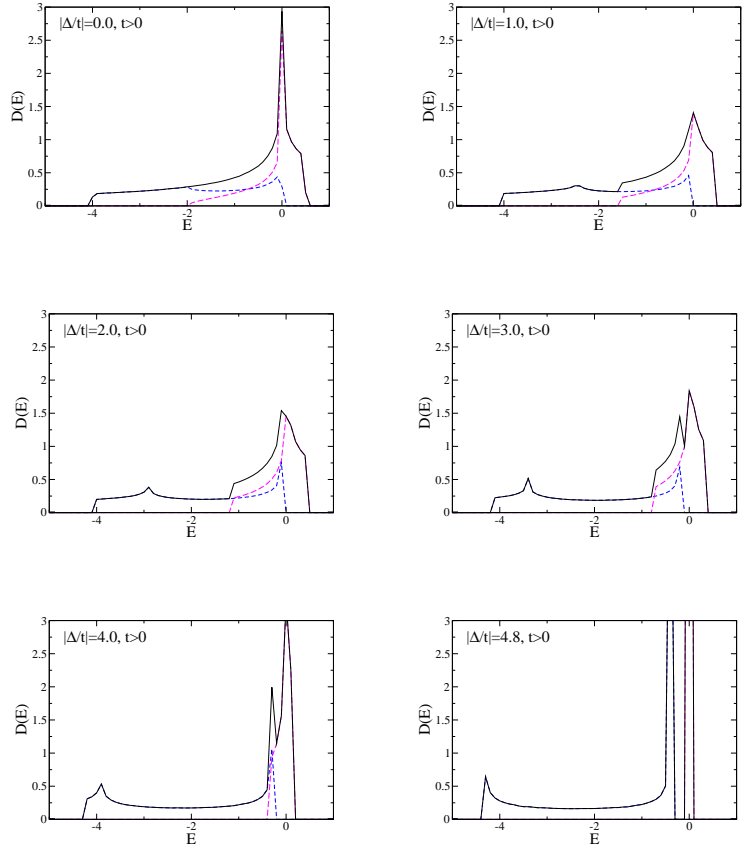

\epsfig{file=DOS_tgt0_Delta_eq0p000.eps,width=4.25cm} \hspace{1cm}
\epsfig{file=DOS_tgt0_Delta_eq1p000.eps,width=4.25cm}\vspace*{1cm}\\
\epsfig{file=DOS_tgt0_Delta_eq2p000.eps,width=4.25cm}\hspace{1cm}
\epsfig{file=DOS_tgt0_Delta_eq3p000.eps,width=4.25cm}\vspace*{1cm}\\
\epsfig{file=DOS_tgt0_Delta_eq4p000.eps,width=4.25cm}\hspace{1cm}
\epsfig{file=DOS_tgt0_Delta_eq4p800.eps,width=4.25cm} \caption{[Color
online] Density of states (in units where $|t|=a=1$) of the fermions for $t>0$. 
 Dashed (blue and pink) lines indicate the
contributions of the individual bands. As in Fig. \ref{fig:DOS-t<0} we see that the antibonding band rapidly narrows at \dt increases. Furthermore, the width of the bonding band is almost independent of \dtn.
Again, the DOS is very different from the non-interacting case, cf. Fig. \ref{dos-gt}, particularly for large \dtn. Also see the online movie.\cite{mov}} \label{fig:DOS-t>0}
\end{figure*}

These conclusions are consistent with changes observed in
the density of states, which we plot in Figs. \ref{fig:DOS-t<0} and \ref{fig:DOS-t>0} (also see the
animations online\cite{mov}), calculated  by treating the fermions as
non-interacting particles. These figures should be contrasted with
the equivalent plots of the non-interacting system (Figs.
\ref{dos-lt} and \ref{dos-gt}). For both signs of $t$ as \dt is
increased the antibonding band narrows into the quasiparticle peak,
familiar from the Mott-Hubbard
transition.\cite{KotliarVollhardtPhysToday} Note that, as the
fermions are hole-like, the states with $E>0$ are filled and the
states with $E<0$ are unoccupied.

\begin{figure}
\epsfig{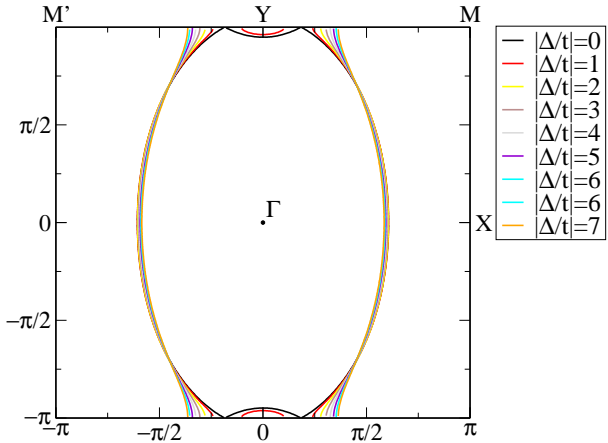} \caption{[Color
online] Fermi surface of the fermions for $t<0$. For small
$|\Delta/t|$ there are small Fermi pockets around the Y points.
Comparing the Fermi surface with the band structure of the fermions
shows that these pockets arise from the bonding band. One therefore
expects that these fermions are lighter than those in the
quasi-one-dimensional sheets, which arise from the antibonding band
(cf. Fig. \ref{fig:cyclo}). The Brillouin zone is that of the ionic
Hubbard model (cf. Fig. \ref{fig:unit-cells}e).
} \label{fig:Fs-t<0}
\end{figure}

\begin{figure}
\epsfig{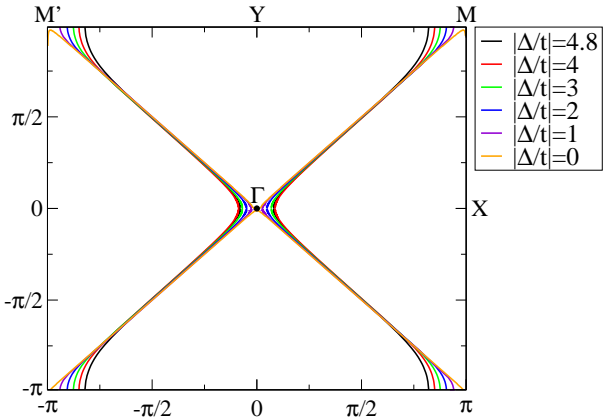} \caption{[Color
online] Fermi surface of the fermions for $t>0$. For $t>0$ the
bonding band is always completely filled. Therefore this Fermi
surface arises entirely from the antibonding band. As \dt is
increased only rather subtle changes are observed in the Fermi
surface, most notably, the nesting is slightly decreased. The
Brillouin zone is that of the ionic Hubbard model (cf. Fig.
\ref{fig:unit-cells}e).} \label{fig:Fs-t>0}
\end{figure}

We plot the the Fermi surfaces of the quasiholes in Figs.
\ref{fig:Fs-t<0} and \ref{fig:Fs-t>0}. Both signs of $t$ exhibit
strong nesting, suggesting that there may be magnetic instabilities,
which we do not consider here. For $t<0$ and $|\Delta/t|<1.5$
there are small two-dimensional Fermi pockets of light
quasiholes arising from the bonding band, as well as
the quasi-one-dimensional sheets arising form the antibonding band.
For $|\Delta/t|\rightarrow0$ the small Fermi pocket occupies $0.7\%$ of the Brillouin zone. As \dt increases the Fermi pocket shrinks and it vanishes at $|\Delta/t|=1.5$.

In the \A materials the unit cells are twice the size of that of our
model. This results in a zone folding of the Brillouin zone in the
$x$-direction, cf. Fig. \ref{fig:unit-cells}. This will result in
the formation of additional small pockets around the $X$ and $M$
points (of the \A Brillouin zone, Fig. \ref{fig:unit-cells}f), which
are formed from fermions in the antibonding band. If one simply
folds our results into the \A Brillouin zone the pockets from the
bonding band occupy 1.4\% of the Brillouin zone for
$|\Delta/t|\rightarrow0$ (as the Brillouin zone has halved in size,
while the pocket area remains unchanged) and the pockets around the
$X$ and $M$ points arising from the antibonding band occupy 8.6\%
and 9.9\% of the Brillouin zone respectively.


\subsection{Spectral density}

The combination of small \dt and large $U/|t|$ significantly changes
the dispersion relations of the quasiparticles from those of the
non-interacting electrons, compare section \ref{sect:bs-quasiholes}
with Appendix \ref{sect:non-int} and particularly Figs.
\ref{fig:bs-t<0} and \ref{fig:bs-t>0} with Figs. \ref{bs-lt} and
\ref{bs-gt}. First, the antibonding band  has a much weaker
dispersion
 than for $U=0$ than for $U=\infty$.
This is seen in ARPES experiments on \Na and \Kn, where the measured Fermi velocity
           is several times smaller than the  value calculated
from the LDA with no sodium ion ordering (cf. Fig. 2i of Ref.
\onlinecite{Qian}). As $\Delta$ increases weight gets shifted from
the antibonding to the bonding band.

\begin{figure*}
$\Delta=0; t<0$ \hspace{3cm} $\Delta=2|t|; t<0$ \hspace{3cm} $\Delta=4|t|; t<0$\\
\epsfig{file=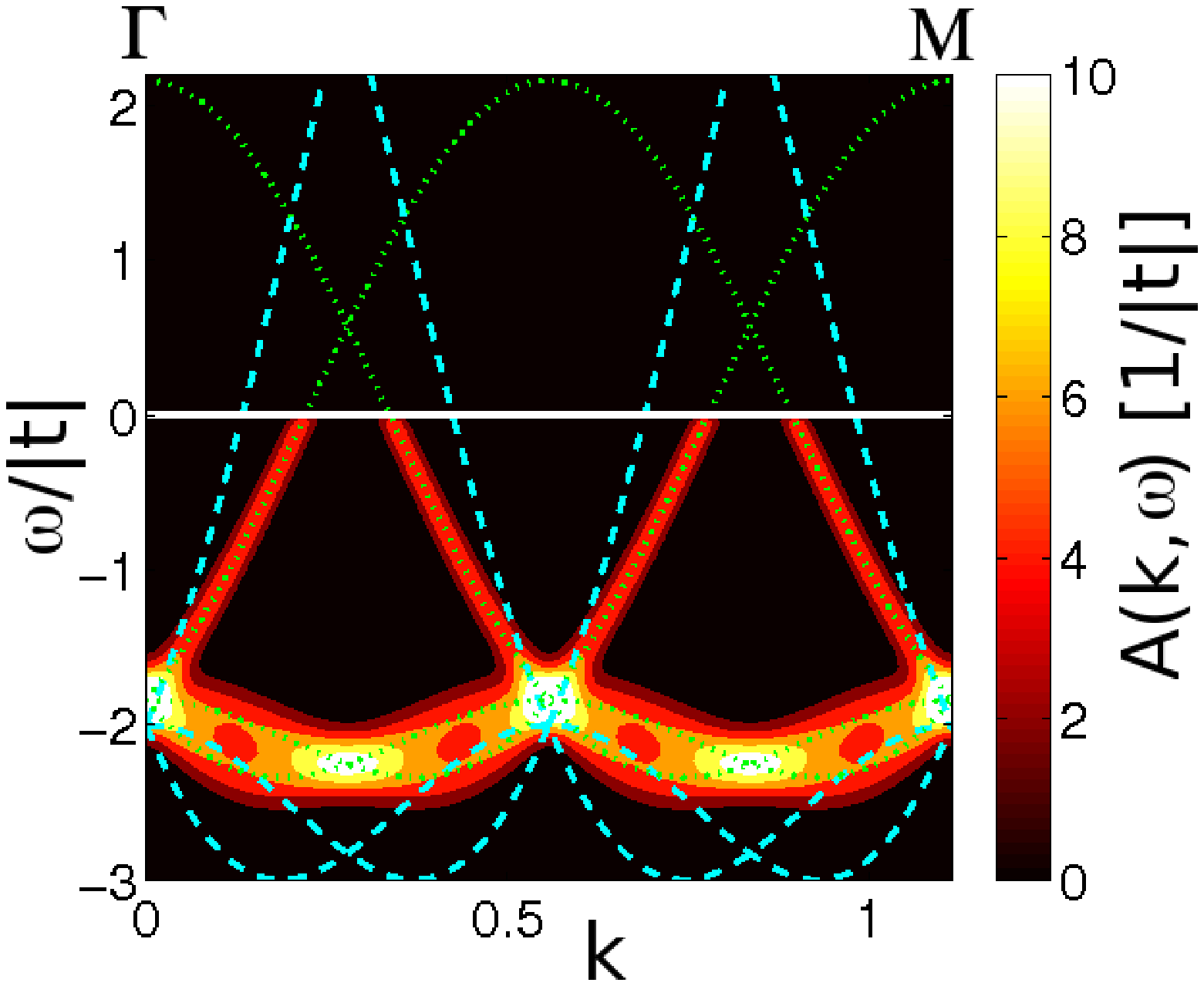,width=5.4cm}
\epsfig{file=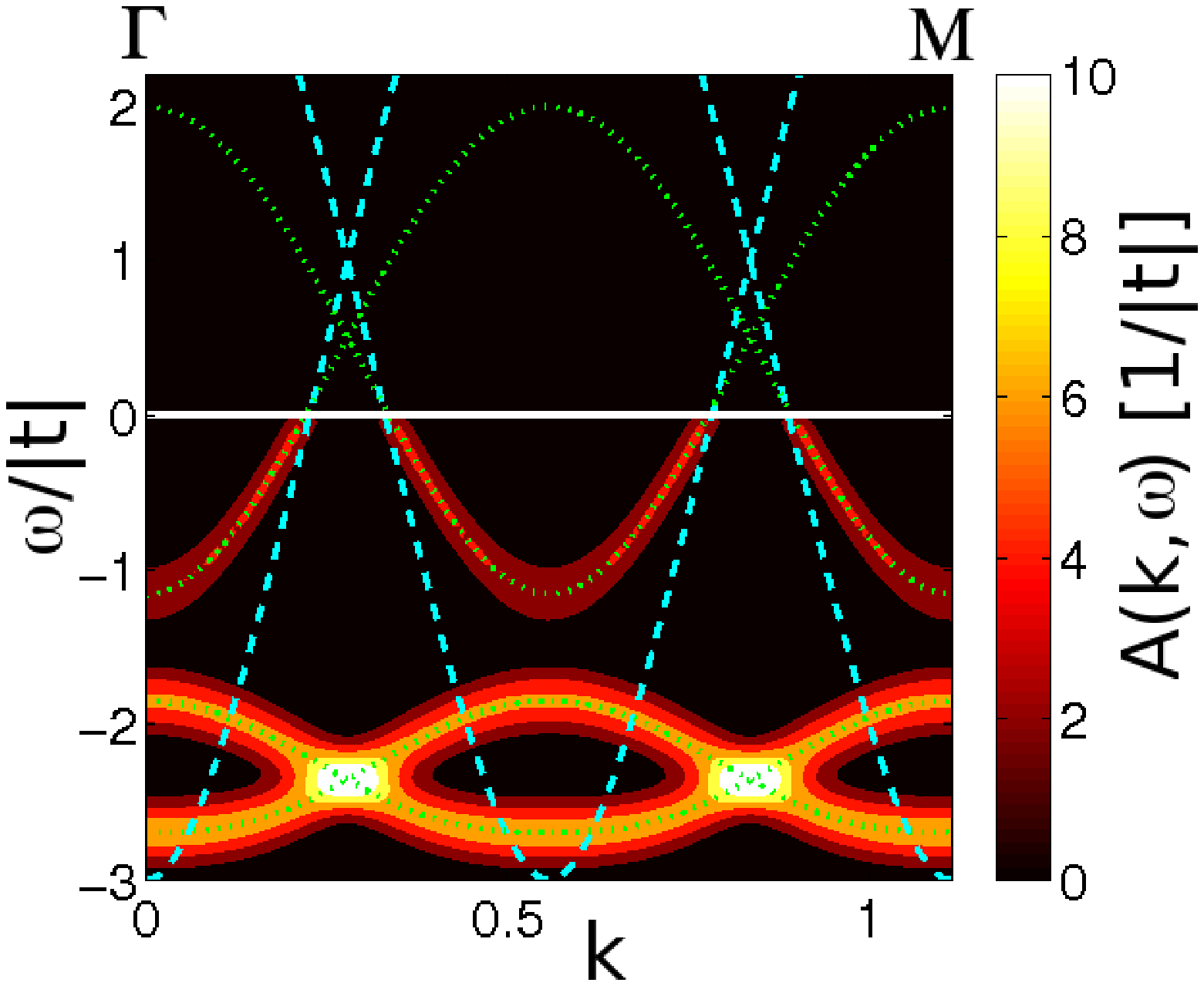,width=5.4cm}
\epsfig{file=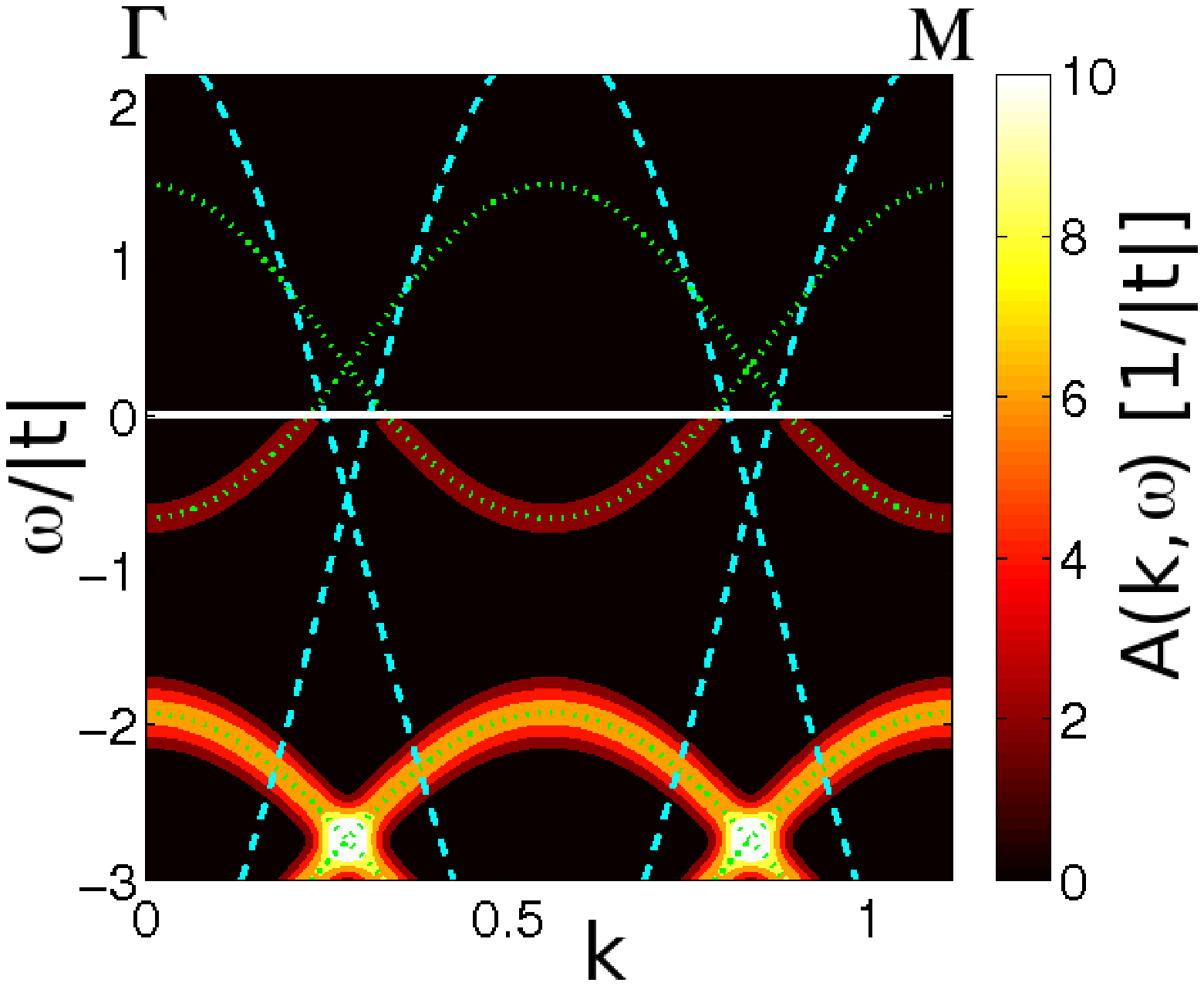,width=5.4cm}\\
$\Delta=6|t|; t<0$ \hspace{3cm} $\Delta=7.9|t|; t<0$\\
\epsfig{file=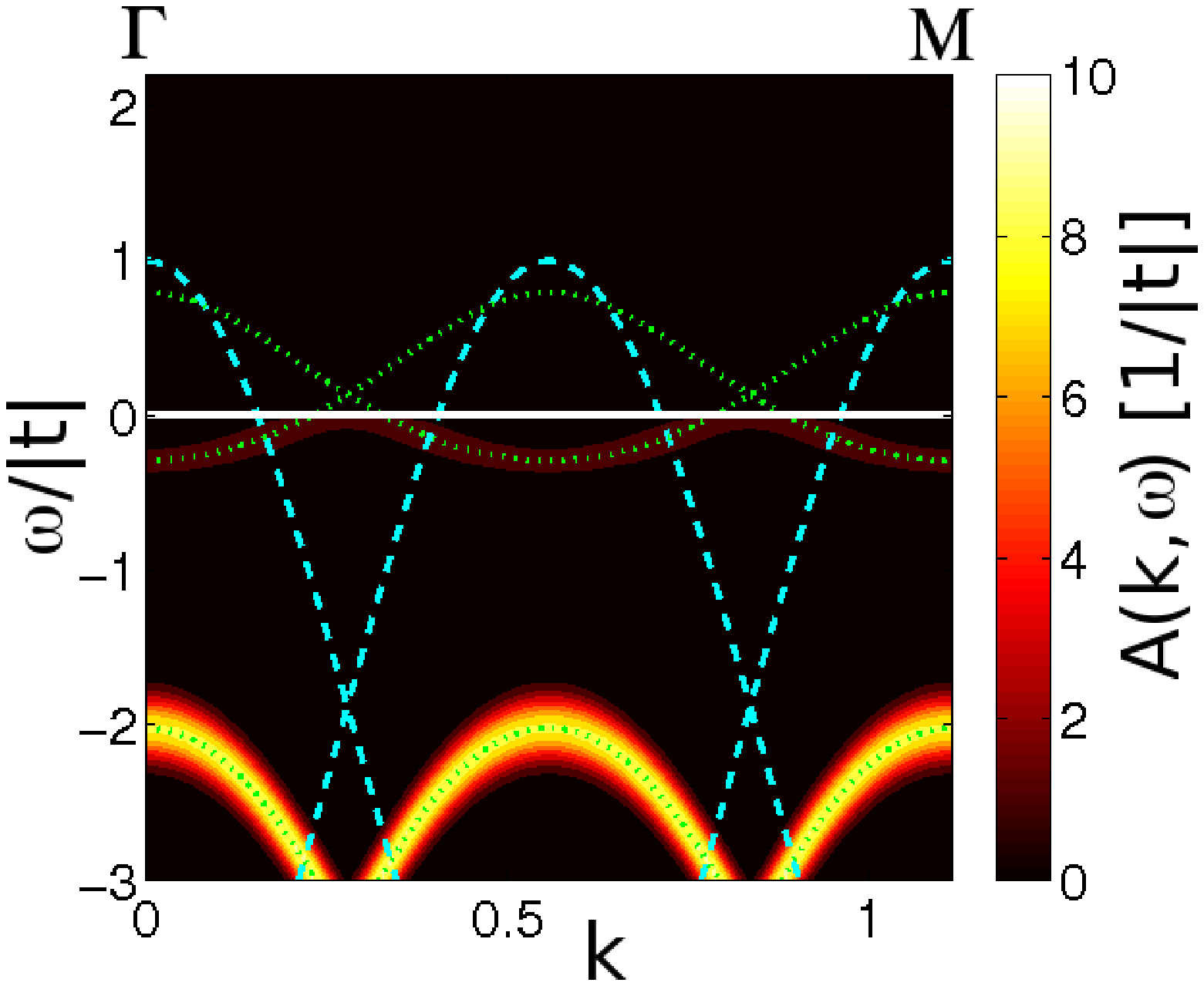,width=5.4cm}
\epsfig{file=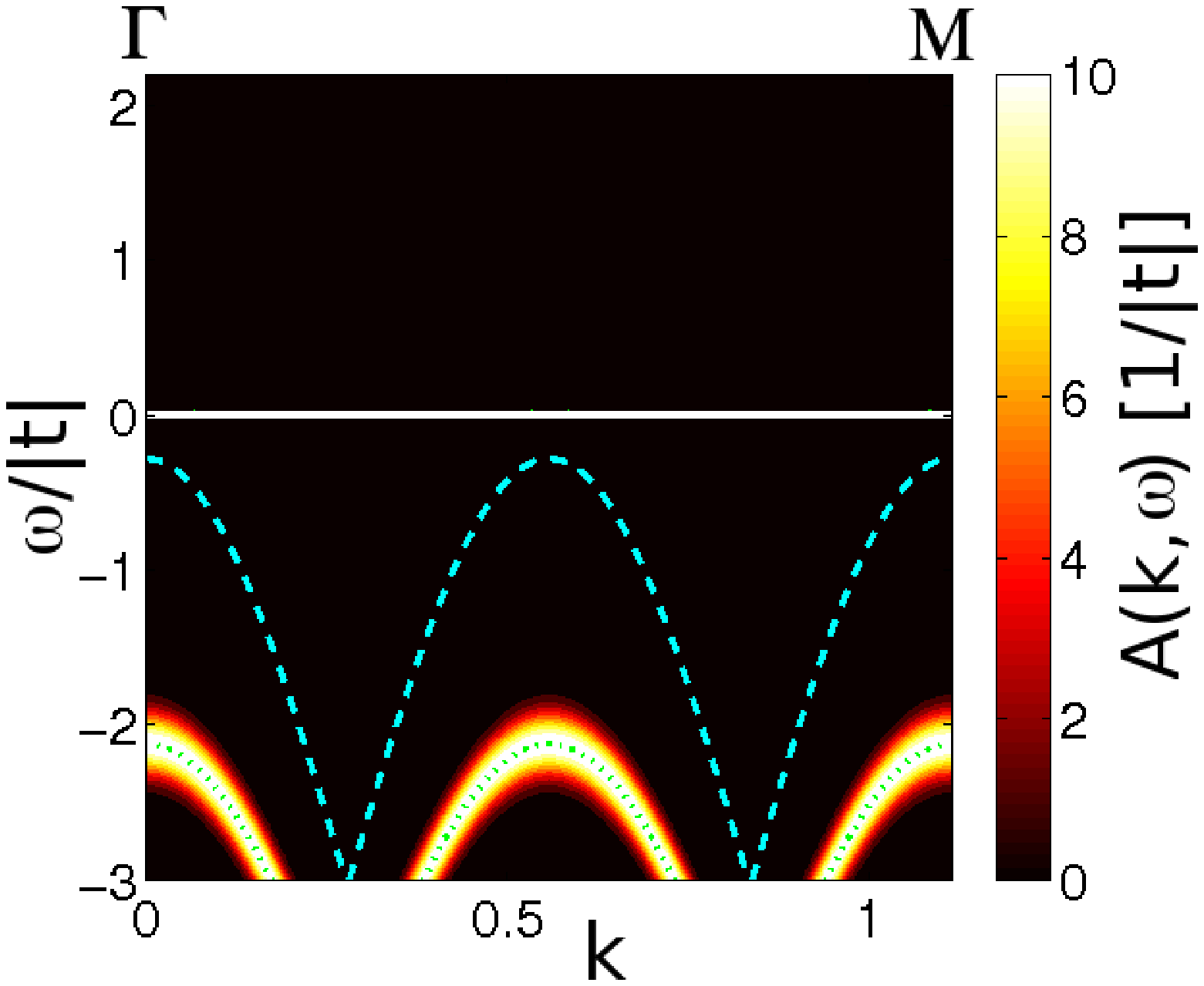,width=5.4cm}
\caption{[Color online] Cut of the spectral density, $A({\bf
k},\omega)$, along the $\Gamma$-M direction of the triangular
lattice Brillouin zone (cf. Fig. \ref{fig:unit-cells}d). $A({\bf
k},\omega)$ is calculated from Eq. (\ref{eqn:ARPES}), broadened as
described in the text below Eq. (\ref{eqn:broadening}), and plotted
for $k_BT=|t|/60\simeq20$~K. Green dotted lines are the slave boson
dispersion, blue dashed lines are the bare dispersion. The abscissa
is in \AA$^{-1}$, with the lattice constant taken as $a=2.82$ \AA,
as is appropriate for \Nan.\cite{Zhang} As \dt is increased the
spectral weight is transferred from the antibonding band to the
bonding band and the antibonding band narrows. Also see the online
movie.\cite{mov}} \label{fig:ARPES-GM}
\end{figure*}

\begin{figure*}
$\Delta=0; t<0$ \hspace{3cm} $\Delta=2|t|; t<0$ \hspace{3cm} $\Delta=4|t|; t<0$\\
\epsfig{file=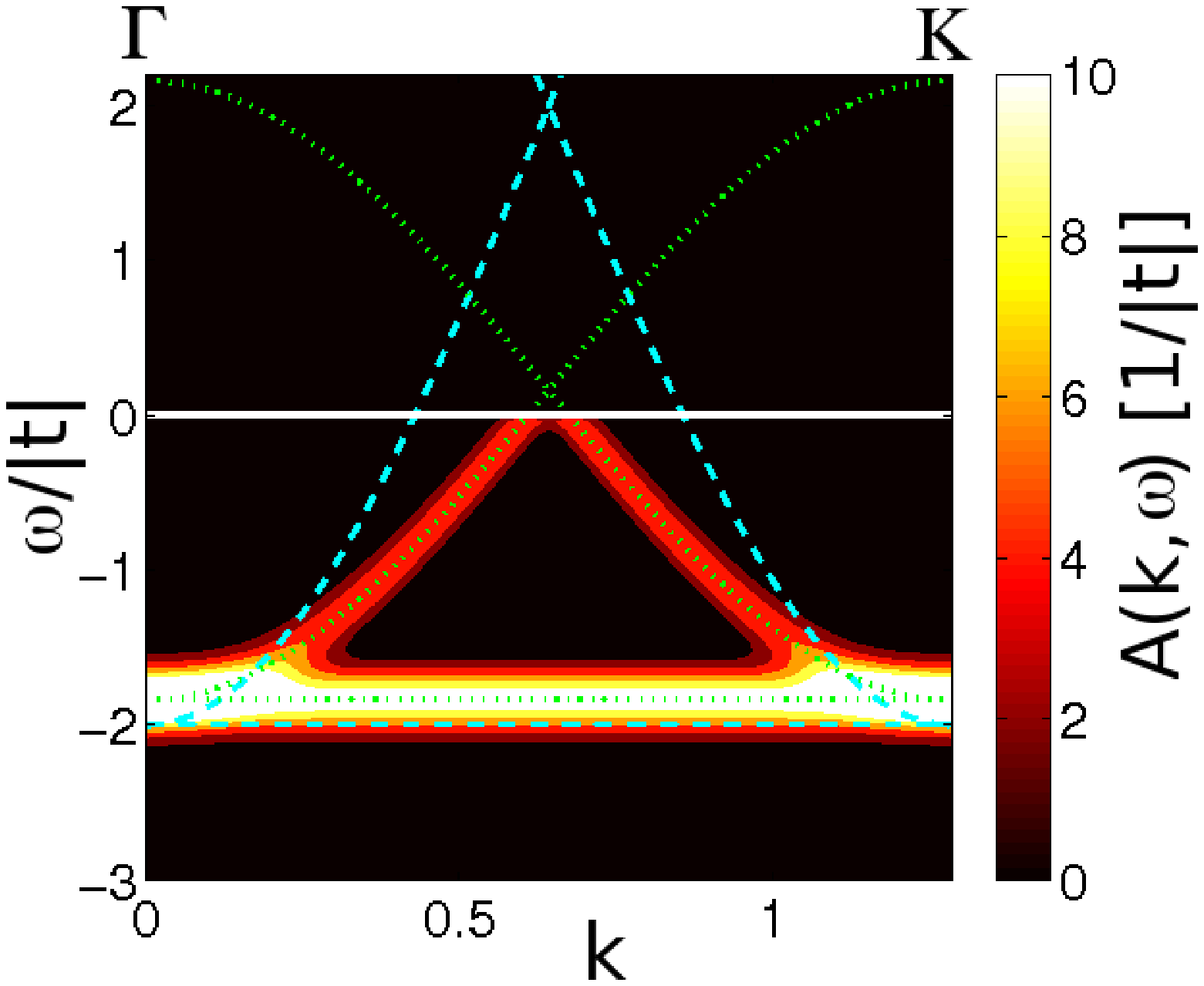,width=5.4cm}
\epsfig{file=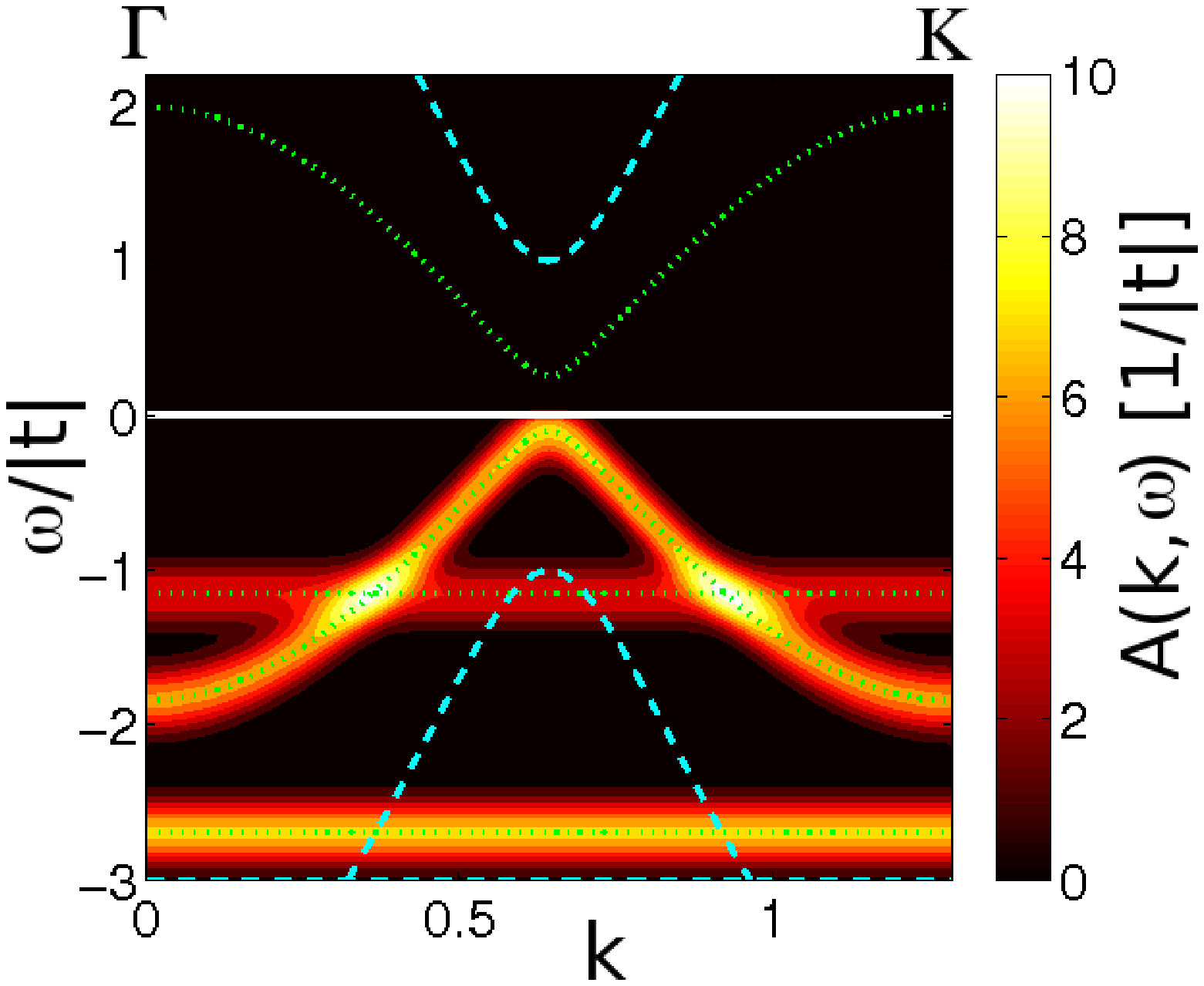,width=5.4cm}
\epsfig{file=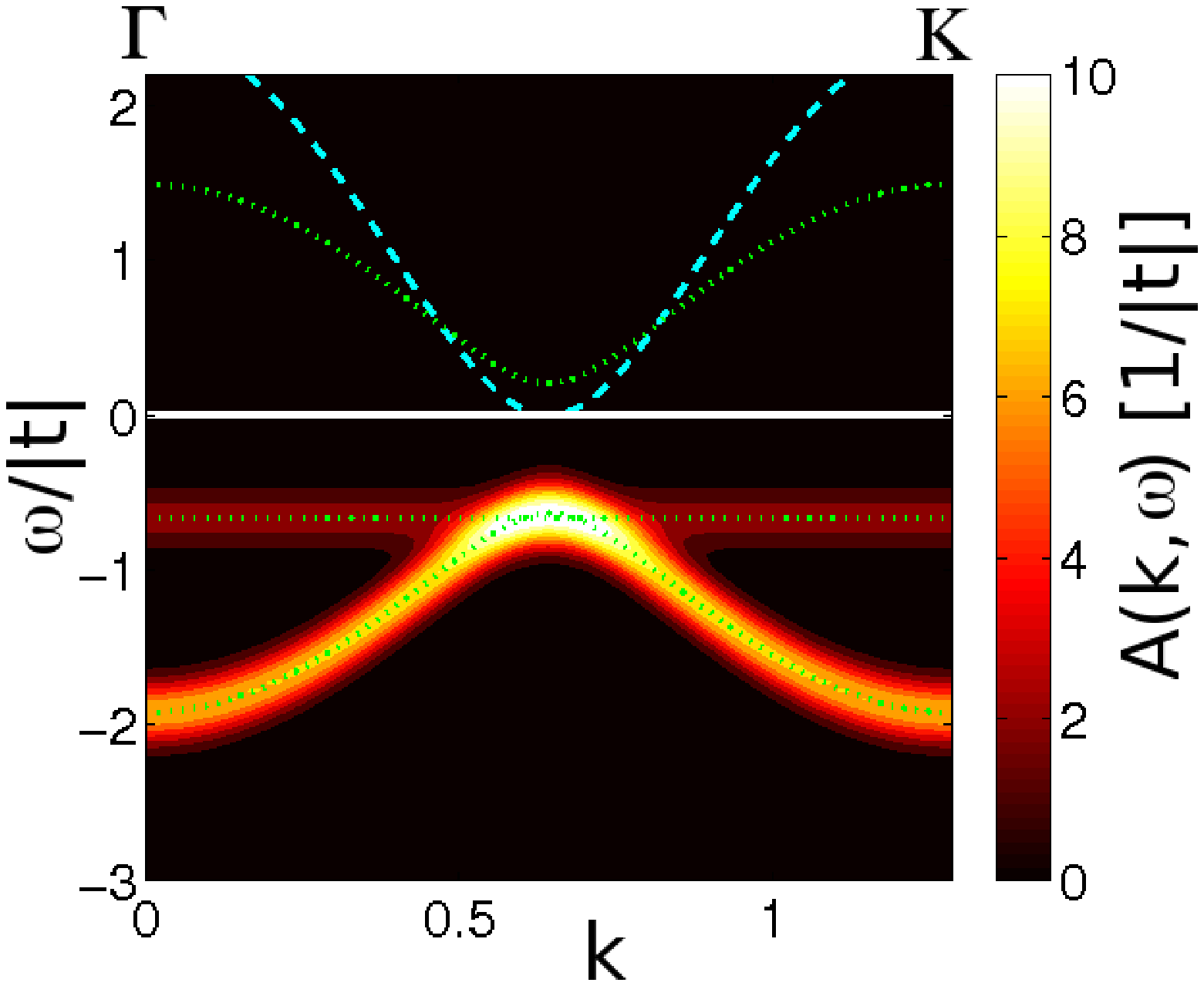,width=5.4cm}\\
$\Delta=6|t|; t<0$ \hspace{3cm} $\Delta=7.9|t|; t<0$\\
\epsfig{file=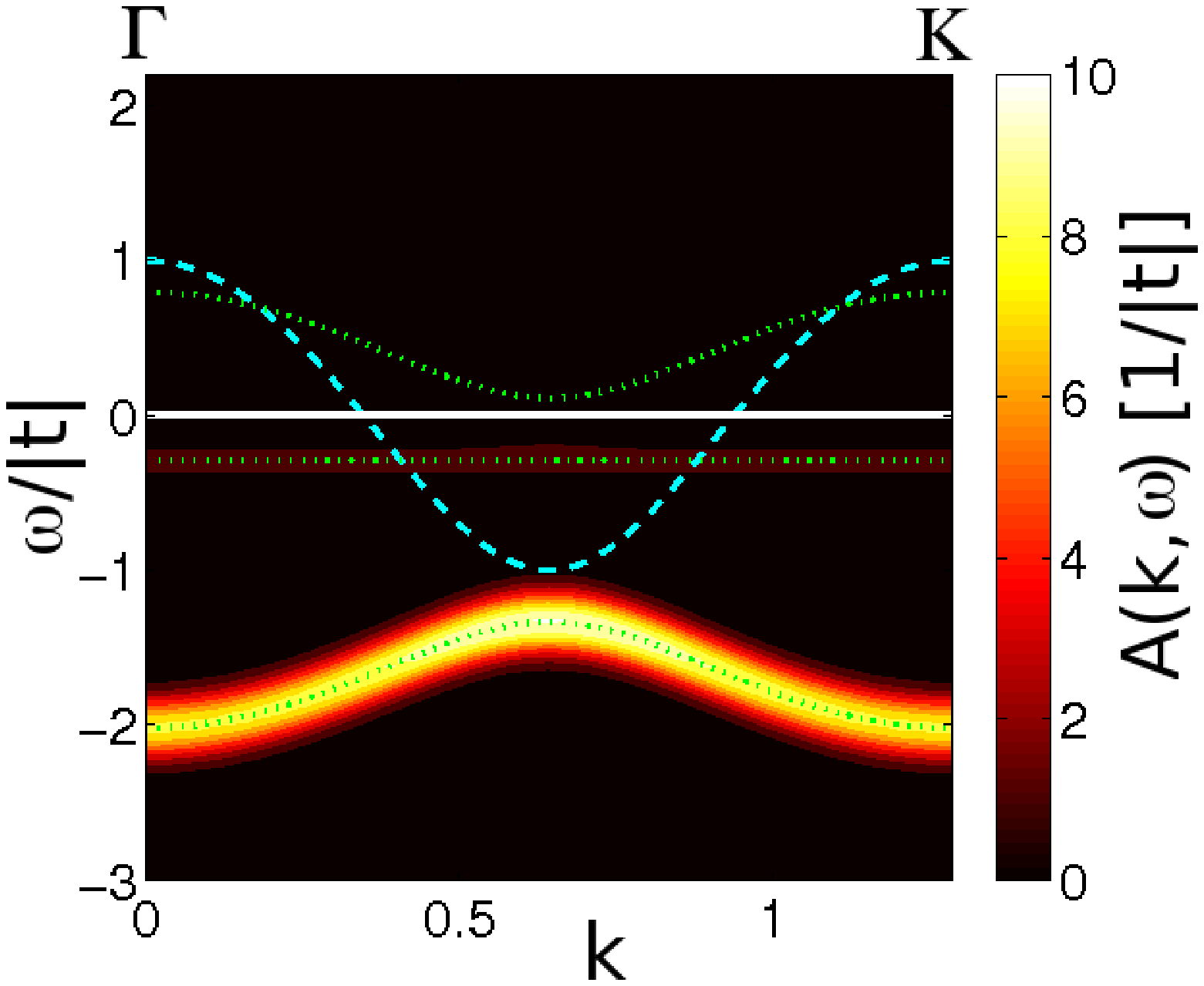,width=5.4cm}
\epsfig{file=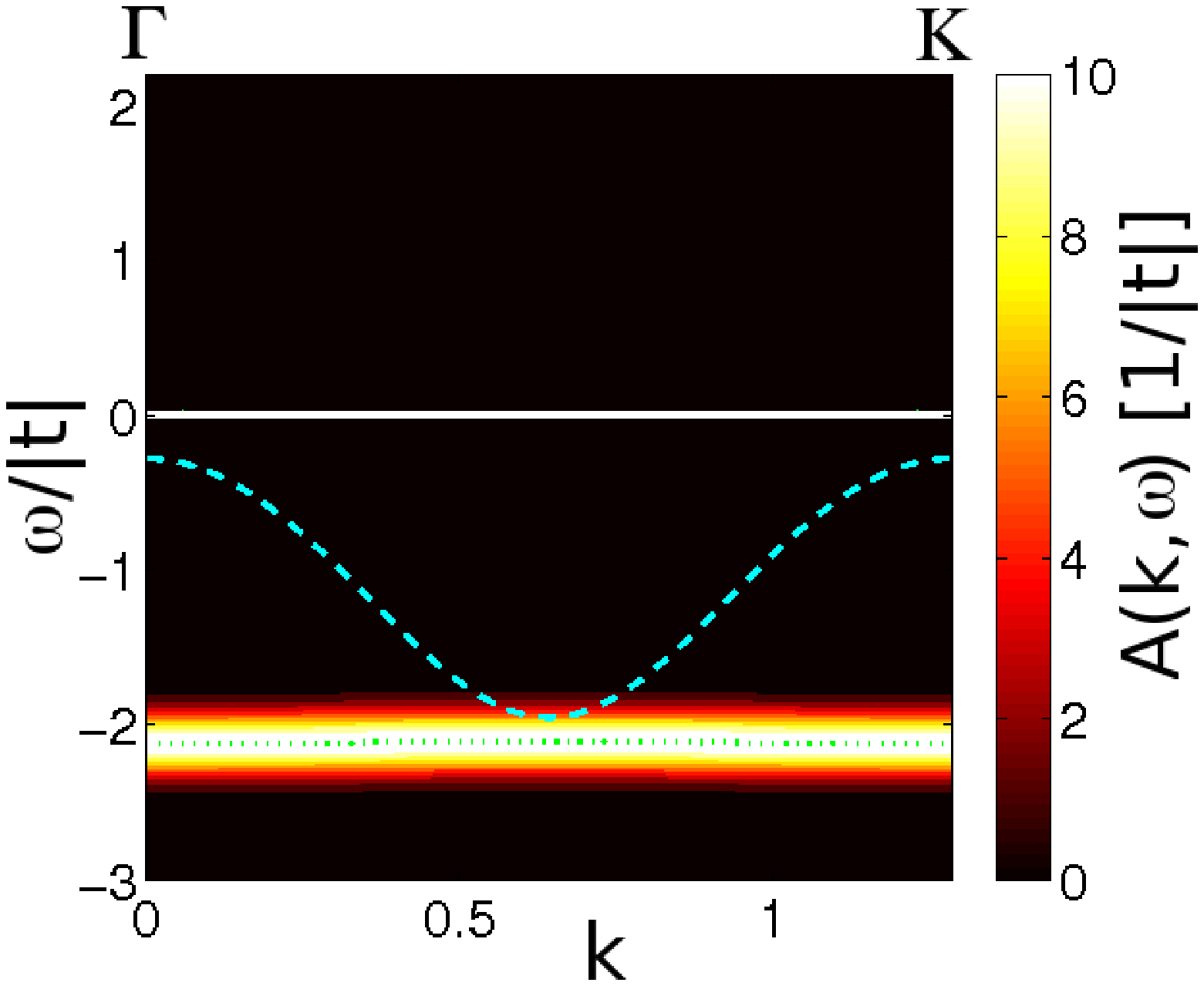,width=5.4cm}
\caption{[Color online] Cut of the spectral density, $A({\bf
k},\omega)$, along the $\Gamma$-K direction of the triangular
lattice Brillouin zone (cf. Fig. \ref{fig:unit-cells}d). $A({\bf
k},\omega)$ is calculated from Eq. (\ref{eqn:ARPES}), broadened as
described in the text below Eq. (\ref{eqn:broadening}), and plotted
for $k_BT=|t|/60\simeq20$~K. Green dotted lines are the slave boson
dispersion, blue dashed lines are the bare dispersion. The abscissa
is in \AA$^{-1}$, with the lattice constant taken as $a=2.82$ \AA,
as is appropriate for \Nan.\cite{Zhang} At small \dt the bonding
band crosses the Fermi energy in this direction (cf. Figs.
\ref{fig:bs-t<0} and \ref{fig:Fs-t<0}). As \dt is increased the
bonding band is pushed below the Fermi energy and spectral weight is
transferred from the antibonding band to the bonding band, which is
not dispersive in this direction. Also see the online
movie.\cite{mov}} \label{fig:ARPES-GK}
\end{figure*}

We plot two different reciprocal space cuts of the calculated
spectral function for representative values of \dt and $t<0$ in
Figs. \ref{fig:ARPES-GM} and \ref{fig:ARPES-GK} (also see the
animations online\cite{mov}). As shown in Fig.
\ref{fig:unit-cells}, the unit cells of $A_{0.5}$CoO$_2$ are twice
as large as the unit cell of Hamiltonian (\ref{ham}). In order to
facilitate comparison with experiment, we have plotted these results
in the Brillouin zone of Na$_{0.5}$CoO$_2$ by `folding over' the
results for Hamiltonian (\ref{ham}) into the Brillouin zone of the
actual materials. At $T=0$ Eq. (\ref{eqn:ARPES}) gives a plane of
$\delta$-functions (in the three dimensional ${\bf k}$, $\omega$
space). It is well known that
\begin{equation}
\delta(\epsilon-\omega)=\lim_{b\rightarrow0}\frac{\exp\left[-{(\epsilon-\omega)^2}/{4b}\right]}{2\sqrt{\pi
b}}.\label{eqn:broadening}
\end{equation}
Therefore we have broadened the results by replacing the
$\delta$-function with a Gaussian with $b=0.01t^2$. We report our
results with the Fermi factor appropriate for
$k_BT=|t|/60\simeq20$~K for $-t=0.1$~eV.

 It can be seen from
Figs. \ref{fig:ARPES-GM} and \ref{fig:ARPES-GK} that as \dt is
increased the antibonding band remains at the Fermi energy, but
loses spectral weight, consistent with the calculated density of
states for fermions (Figs. \ref{fig:DOS-t<0} and \ref{fig:DOS-t>0}).
Meanwhile, the bonding band is shifted down in energy. In the
$\Gamma$-M cut (Fig. \ref{fig:ARPES-GM}) we see that the antibonding
band narrows and loses intensity as \dt is increased. In the
$\Gamma$-K cut (Fig. \ref{fig:ARPES-GK}) we see that the bonding
band moves from the Fermi level at $\Delta=0$ to lower energies such
that it is completely filled at larger values of \dtn. At first
sight Fig. \ref{fig:ARPES-GK} might appear to suggest that the
bonding band also narrows as \dt increases, however it is clear from
Fig. \ref{fig:ARPES-GM} that this is not the case. Rather, as \dt
increases the system becomes more one dimensional (for $t/\Delta=0$,
the system consists of uncoupled one-dimensional chains), increasing
the range of $E_{\bf k}$ in the cuts shown in Fig.
\ref{fig:ARPES-GM} and decreasing the range of $E_{\bf k}$ in the
cuts shown in Fig. \ref{fig:ARPES-GK} (cf. Figs. \ref{fig:bs-t<0}
and \ref{fig:Fs-t<0}).

\subsection{Quantum oscillations}

\begin{figure}
\epsfig{file=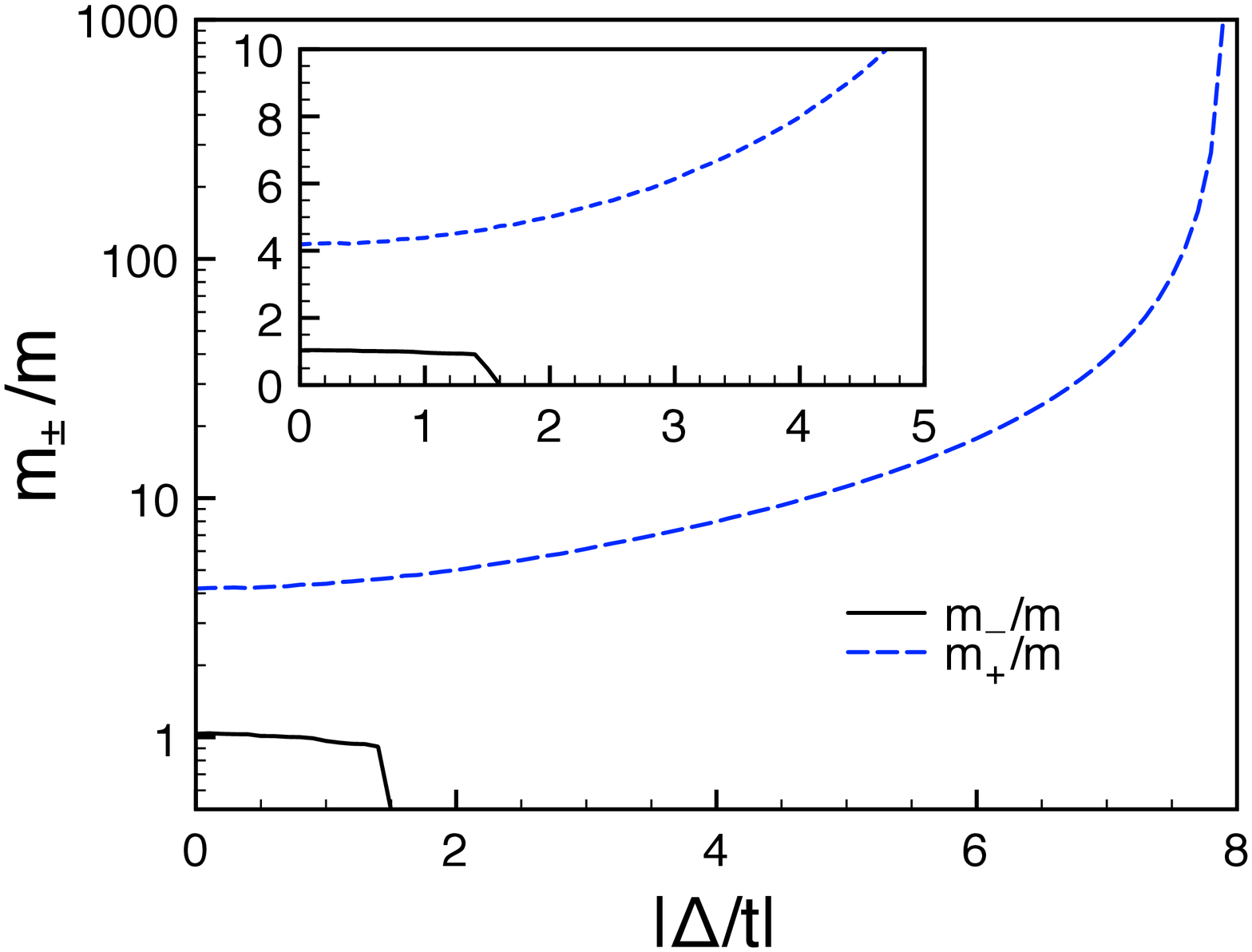,width=8cm} \caption{[Color online]
Variation of the renormalised cyclotron effective mass, $m_\pm$, in
units of the free electron mass, $m$, with \dt for $t<0$. Note that
in the main panel the ordinate is plotted on a logarithmic scale:
the same data is plotted on a linear scale in the inset.
$m_-/m\sim1$ and $m_+/m\sim4$ for small \dtn. The cyclotron
effective mass of the antibonding band diverges as the metal-charge
transfer insulator transition is approached, consistent with our
finding that the metal-charge ordered insulator transition is second
order, cf. Fig. \ref{fig:F-t<0}. The cyclotron effective masses are
calculated from Eq. (\ref{cyclotron-mass}) with parameters
appropriate for \Nan: $t=-0.1$ eV, an in-plane lattice constant,
$a=2.82$ \AA, and a unit cell basal area of $\sqrt{3}a\times 2a$.}
\label{fig:cyclo}
\end{figure}

As we are considering a quasi-two-dimensional metal the cyclotron
effective mass, $m_{\pm}$, associated with Shubnikov-de Haas
oscillations of the part of the Fermi surface arising from the
$\pm$ band is\cite{merino00}
\begin{equation}
m_{\pm} = 2\pi \hbar^2 D_{\pm}(E_F), \label{cyclotron-mass}
\end{equation}
where $D_{\pm}(E_F)$ is the renormalised density of states per spin
at the Fermi energy, $E_F$, arising from the $\pm$ band. We plot the
variation of the cyclotron mass with \dt for parameters relevant to
\Nan, i.e., $t=-0.1$ eV, a basal lattice constant, $a=2.82$ \AA, and
a unit cell basal area of $\sqrt{3}a\times 2a$, in Fig.
\ref{fig:cyclo}. The  cyclotron mass for the bonding band
associated with the small pockets is about one quarter of that for
heavy electrons associated with the antibonding band.

In order to observe the states arising from the antibonding band,
i.e., the quasi-1D sheets on the Fermi surface, in quantum
oscillation experiments a field that is sufficiently large to cause
magnetic breakdown between the two bands must be applied. The field
required for magnetic breakdown and the large effective mass
associated the antibonding band make it more difficult to observe the breakdown
orbit than than the small pockets in quantum oscillation experiments.

For realistic parameters for \Nan, $\Delta\sim |t|\ll U$ and $t<0$
(cf. section \ref{sect:NaTheory}), the slave boson theory presented
above suggests that the system is metallic. However, an important
question, which we cannot conclusively answer because of  the
uncertainty in the precise value of \dt relevant to \Nan, is: is \dt
large enough in \Na that the bonding band is completely filled? For
$|\Delta/t|<1.5$ there are two features in the Fermi surface (cf.
Fig. \ref{fig:Fs-t<0}): small hole-like pockets of light fermions
arising from the bonding (nearly filled) band, and
quasi-one-dimensional sheets of heavier electron-like
quasiparticles. This seems to be quite consistent with quantum
oscillations,\cite{Balicas} which are observed to have a frequency
correspond to orbits that encompass less than 1\% of the first
Brillouin zone, in apparent violation of Luttinger's theorem. A
natural explanation of this experiment is that Balicas
\etal\cite{Balicas} are only able to observe the, small, light,
quasiparticles in the bonding band and do not see oscillations
arising from the antibonding band, either because the quasiparticles
in the antibonding band are two heavy, or because the field they
applied is insufficient to cause magnetic breakdown of the `gap'
between the quasi-one-dimensional sheets and the two-dimensional
pockets. However, for $|\Delta/t|>1.5$ the small Fermi pockets are
absent as the bonding band is filled, and our results would be
inconsistent with the quantum oscillations observed by Balicas \etal

\subsection{A picture of \Nan}

The experimental situation, particularly in Na$_{0.5}$CoO$_2$, is
somewhat confusing. Several experiments suggest an insulating state:
resistivity versus temperature has a negative gradient at low
temperatures,\cite{Foo} and angle resolved photoemission spectroscopy
(ARPES)\cite{Qian} and optical conductivity\cite{wang} both observe
a gap, all of which suggests that below 51 K Na$_{0.5}$CoO$_2$ is a
insulator. However, quantum oscillations are observed,\cite{Balicas}
which suggest that there are quasiparticles and that the system is
metallic with a small Fermi surface (occupying less that 1\% of the
cross-sectional area of the first Brillouin zone) and an effective
mass similar to the free electron mass. 

Below we propose a picture,
inspired by both the mean field slave boson theory (above) and our
previous exact diagonalisation calculations for the same ionic
Hubbard model,\cite{JaimeLetter,JaimeLong} that offers a possible
reconciliation of these seemly contradictory experiments.
However, it is important to stress that this is a tentative proposal and does not correspond to the solution found in either theory. Nevertheless, we will argue, below, that this picture is plausible given the limitations of the mean field slave boson theory and the finite size effects in the exact diagonalization calculations. Further calculations will be required to discover whether this proposal is indeed supported  by the theory of the ionic Hubbard model.

The fundamental definition of a metal is a substance with charge
carrying  excitations at arbitrarily low energies above the ground
state. In most situations it follows that the resistivity
monotonically increases with temperature as increasing the
temperature decreases the quasiparticle lifetime. However, this is
not necessarily the case. For example, in many strongly correlated
systems, such as the organic superconductors,\cite{JPCMreview} the
charge transport is incoherent  at high temperatures - this leads to
a resistivity with a broad maximum slightly above the temperature
where quasiparticle coherence is lost.\cite{MerinoDMFT} Above, we
have shown that the slave boson theory of Hamiltonian (\ref{ham})
produces two bands. We will assume some features of these bands survive even in the exact solution even if the simple single particle states, of the slave boson theory, do not. The antibonding band is narrow and it is likely
that coherent quasiparticles do not emerge from these states until
quite low temperatures. Within slave boson theories the coherence
temperature can  be estimated from the Bose condensation
temperature, $T_B$ of the slave bosons,\cite{LNW} which, in 2D, is
proportional to the density of bosons, $\rho_B$. For
$|\Delta/t|\rightarrow0$ there are 0.993 quasiholes per unit cell in
the antibonding band, corresponding to $\rho_B=0.007$, and the
density of bosons decreases as \dt is increased, going to zero at
\dtn=1.5. Estimating the Bose condensation temperature  (cf. Refs.
\onlinecite{LNW} and \onlinecite{Lee92}) by either
$T_{B}\approx(2\pi \rho_B)/(\sqrt{3}a^2m_B)$, where $a$ is the basal
plane lattice constant (2.82 \AA~ for \Nan) and $m_B\approx2m$ is
the mass of the boson (which is the simplest approximation given
that the boson corresponds to two electrons), or $T_B\approx 4\pi
\rho_Bt$ yields $T_B\sim100$ K for $\rho_B=0.007$. However it is
well known that these simple formulae dramatically overestimate the
coherence temperature as they neglect gauge
fluctuations.\cite{LNW,Lee92} Thus the coherence temperature ($T^*\ll100$~K) is expected to be extremely low in the experimentally
relevant parameter regime.

 This low coherence temperature is consistent with the observed incoherent transport evident from the
resistivity above $\sim$51 K. As the temperature is lowered below 51
K the resistivity rises rapidly, suggesting that a gap opens. The
slave boson calculations do not predict a true gap unless
$|\Delta/t|\gg1$. Experiments observe magnetic
ordering,\cite{bobroff,schulze,gasparovic,yokoi} which we have not
considered here, below 88 K, but, as the gap does not appear to open
until 51 K, one cannot conclude that the magnetic order causes the
gap to open. However, previous exact diagonalisation
calculations\cite{JaimeLetter,JaimeLong} suggest that in the
experimentally relevant parameter regime, $-t\sim\Delta\ll U$,
non-local correlations, not captured by our mean field treatment,
drive the formation of a covalent insulator. We propose that a gap opens on the
antibonding band at 51 K due to non-local correlations, analogous to
the covalent insulator. But, we also propose that the bonding band
remains ungapped.

This proposal is rather similar to the theory of the orbital-selective Mott transition. The question of whether,  in multi-band systems, one band can becoming insulating, due to electronic correlations, while others remain metallic has been extensively discussed recently,\cite{OSMT} motivated, in part, by the volume collapse in Ce (Ref. \onlinecite{Held}), the heavy fermion behaviour of Ca$_{2-x}$Sr$_x$RuO$_4$ (Ref. \onlinecite{Anisimov}) and the nodal--anti-nodal differentiation in the cuprates.\cite{Ferrero}  Although the debate is ongoing, the current theoretical consensus is that orbital-selective Mott transitions can occur for some parameter regimes of some models.\cite{OSMT} Therefore, testing our proposed picture will require calculations beyond the mean-field slave-boson theory presented above.

Given the important differences between the current mean field
calculations and our previous exact diagonalisation calculations,
particularly the critical value of \dt for the metal-insulator
transition, it is important to ask, whether this explanation is
likely to hold in the exact solution and, more importantly, for the real
material. Firstly, we note that the real space charge
disproportionation calculated from our mean field theory is in very
good agreement with exact diagonalisation calculations in the
experimentally relevant regime. This gives one hope that the filling
of the bands calculated from the mean field theory may also be
reasonable (although non-local Coulomb correlations could cause additional hybridisation effects, cf. Ref. \onlinecite{JaimeLong}). Secondly, we note that the finite size clusters that
were, necessarily, studied in the exact diagonalisation calculations
mean that it is extremely difficult to accurately study the bonding
band. Taking a, typical, 18 site (9 unit cell) cluster, a density of
0.007 holes per unit cell means that one would expect less than
1/16$^\textrm{th}$ of a hole in the bonding band in the exact
diagonalisation calculation. However, exact diagonalisation
calculations do accurately reproduce the band filling of the analytical exact
solution for $U=0$.\cite{JaimeLetter,JaimeLong} Therefore, it is not clear, at present, whether
the exact diagonalisation calculations can resolve the tiny hole
pockets suggested by our calculations and the quantum oscillation
experiments.

As the exact diagonalisation calculations do not rule out the
possibility that small weakly correlated Fermi pockets containing a
tiny number of charge carries remain even in the low temperature
`insulating' state, let us briefly discuss how this would produce a
consistent explanation of the experiments on \Nan.

It is important to note that the two bands act as parallel channels
for charge transport. Above 51 K charge transport in the antibonding
band will be incoherent (assuming the coherence temperature $<50$ K,
as we have argued above) and the resistivity in this channel can be
estimated to be $\gtrsim\hbar c/e^2$, where $c$ is the interlayer
spacing, i.e., at or above the Mott-Ioffe-Regel limit. Below 51 K a
gap opens on the antibonding band, but as the gap is small, ${\cal
O}(t)$ (Ref. \onlinecite{JaimeLetter,JaimeLong}) the resistivity in
the antibonding channel is not increased by orders of magnitude from
the resistivity value in the incoherent metallic state. However,
because of the extremely low charge carrier concentration in the
bonding band (metallic channel) the contribution of the bonding band
to the conductivity will be even smaller than that of the
antibonding band (incoherent/insulating channel). Therefore the
metallic channel will be unable to `short circuit' the
incoherent/insulating channel. This contention is supported by the
fact that even above 51 K the resistivity is larger than the
Mott-Ioffe-Regel limit. Thus the observation that the resistivity
decreases with increasing temperature is not necessarily
inconsistent with the observation of quasiparticles in quantum
oscillation experiments.

Foo \etal\cite{Foo} found that the conductivity, $\sigma$, of
Na$_{0.68}$CoO$_2$ is very similar to that of Na$_{0.31}$CoO$_2$ and
that both of these are an order of magnitude larger than the
conductivity of \Nan. As $\sigma\propto n$, one expects that the
contribution of the electrons in the bonding band to the
conductivity of \Na will be two orders of magnitude smaller than the
total conductivity of Na$_{0.68}$CoO$_2$ or Na$_{0.31}$CoO$_2$.
Therefore Foo \etaln's results are consistent with the,
counterintuitive, picture that the insulating band short-circuits
the metallic band!



ARPES, which measures $A({\bf k},\omega)$, has been reported in the
directions $\Gamma$-M and $\Gamma$-K, i.e., in the directions of the
cuts shown in Figs. \ref{fig:ARPES-GM} and \ref{fig:ARPES-GK}
respectively. However, our slave boson calculations do not capture
the gap on the antibonding band required in our proposal. Therefore
the gap observed in the ARPES is likely to correspond to the
`covalent insulator' gap on the antibonding band.
Optical conductivity does not observe a Drude peak,
however the height of the Drude peak is proportional to the charge
carrier density, which would make the tiny Fermi pockets suggested
by our slave boson calculations difficult to observe.

As one expects that the Fermi pockets do not contribute significantly to either the resistivity or the Hall coefficient the correspondence of our mean-field slave boson theory to the empirical low energy Hamiltonian studied by Choy \etal\cite{Phillips} (discussed in section \ref{sect:Choy}) become important. The non-local correlations on the antibonding band open up a gap on the quasi-one-dimensional sheets of the Fermi surface. This provides the gap assumed by Choy \etal and suggests that their results will carry across to our model. This suggests that the measured temperature dependence of both the resistivity and the Hall coefficient in \Na have a natural explanation in terms of the ionic Hubbard model. However, clearly further calculations are require to test the validity of this \emph{ad hoc} proposal and its relevance to \Nan.

\section{Summary and Conclusions}\label{sect:conc}

We have presented a mean-field slave boson theory of the ionic
Hubbard model on the triangular lattice with alternating stripes of
site energy. This model has two bands: one of which is weakly
correlated and nearly filled or filled, the other is nearly
half-filled or half-filled and hence strongly correlated. The
results depend strongly on the sign of $t$. The light band is always
filled for $t>0$, but only becomes filled at $|\Delta/t|=1.5$ for
$t<0$. A metal-charge transfer insulator transition occurs at larger
\dt (5.0 for $t>0$ and 8.0 for $t<0$), when complete charge
disproportionation occurs and one sublattice is filled and the other
in half filled.

We have also proposed a speculative picture of \Nan. In particular,
we have argued that the observed quantum oscillations arise from
quasiparticles in the bonding band, but these are not seen in the
zero field resistivity and several other experiments because the low density of charge carriers
in the bonding band mean that the metallic bonding band is not able
to `short circuit' the incoherent/insulating antibonding band. Calculations beyond the mean field slave boson calculations, which we  have presented above, will be required to test this picture.

Another important avenue for future work will be to allow for magnetic
ordering in the theory as this may improve the degree of agreement with
experiments and allow one to compare with other experiments such as
NMR\cite{bobroff,schulze} and neutron
scattering.\cite{gasparovic,yokoi}

\section*{Acknowledgements}

We thank H. Alloul, J. Bobroff, M.-H. Julien, and R. R. P. Singh for helpful discussions. Numerical calculations were performed on the Australian Partnership for
Advanced Computing national facility under the a grant from the
merit allocation scheme. This work was supported by the Australian
Research Council (ARC) under the discovery program (project numbers
DP0557532 and DP0878523), MEC (project CTQ2005-09385-
C03-03) and MICINN (project CTQ2008-06720-C02-02). B. J.
P. was supported by the ARC under the Queen Elizabeth II scheme (DP0878523). J. M.
acknowledges financial support from the Ram\'on y Cajal program. R.H.M. was
supported by the ARC under the APF scheme (DP0877875).

\appendix

\section{Non-interacting electrons}\label{sect:non-int}



\begin{figure}
\begin{center}
\epsfig{file=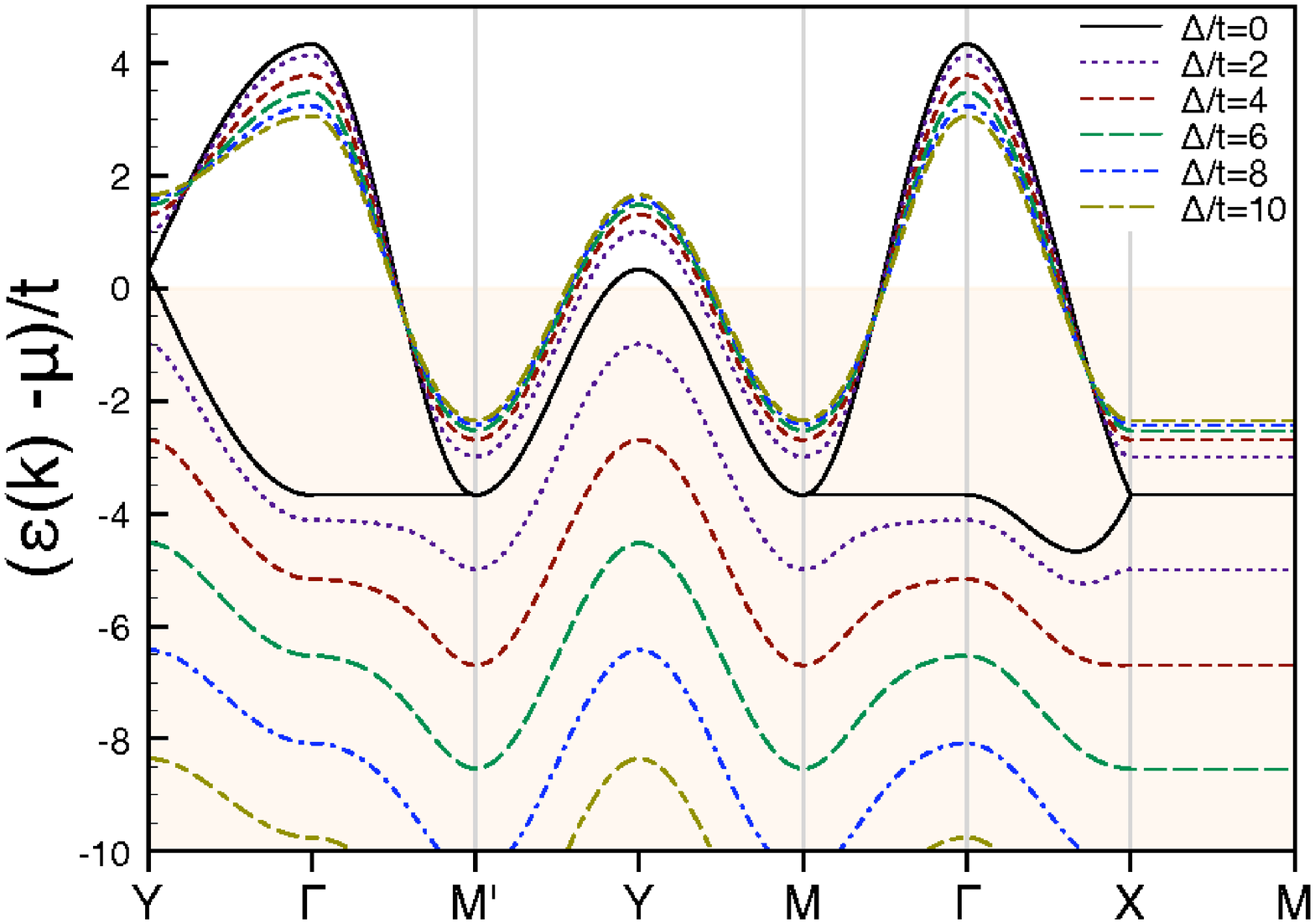,width=8.5cm,angle=0}\caption{Band
structure of the ionic Hubbard model for $U=0$, at three-quarters
filling, and $t<0$. Increasing \dt makes the system increasingly
anisotropic, which serves to flatten the bands slightly in some
directions. The constraint of three quarters filling `pins' the
antibonding band to the Fermi energy, but increasing \dt pushes the bonding
band to lower energies.} \label{bs-lt}
\end{center}
\end{figure}

\begin{figure}
\begin{center}
\epsfig{file=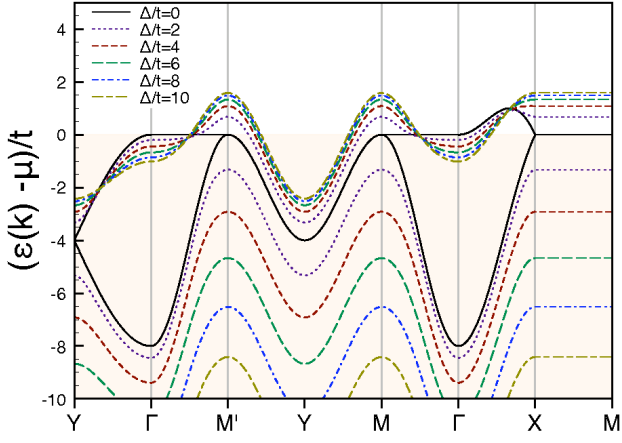,width=8.5cm,angle=0} \caption{Band
structure of the ionic Hubbard model for $U=0$, at three-quarters
filling, and $t>0$. This band structure is markedly different from
the $t<0$ case (Fig. \ref{bs-lt}). This is consistent with the
general expectation that  particle-hole symmetry is absent on
frustrated lattices and hence that the sign of $t$ has strong effects on
the physics.} \label{bs-gt}
\end{center}
\end{figure}

For $U=0$ there are two bands,  bonding (-) and antibonding
(+), 
and whose dispersion relations are
\begin{eqnarray}
\varepsilon_{\bf
k}^\pm&=&-2t\cos(k_x) \notag\\&&
\pm\sqrt{\Delta^2/4+(4t\cos(k_x/2)\cos(k_y/2))^2}, \label{eqn:dispersion}
\end{eqnarray}
where $k_x$ and $k_y$ are defined in the reduced ($1\times\sqrt{3}$) Brillouin zone of Hamiltonian (\ref{ham}), shown in Fig. \ref{fig:unit-cells}e.
These dispersion relations are plotted in Figs.  \ref{bs-lt} and \ref{bs-gt}
for $t<0$ and $t>0$ respectively.

To calculate where the bonding band becomes filled we need to
consider the maximum of $\varepsilon_{\bf k}^-$. This is
complicated by the fact the chemical potential, $\mu$, is a function of $\Delta$. For $t>0$ the
top of the bonding band is at
$\varepsilon_X^-=\varepsilon_M^-=\varepsilon_{(\pi,y)}^-$. Thus we
find that the bonding band become full at $\Delta\geq\Delta^>\equiv2\mu(\Delta^>)+4|t|$.
However, for $\Delta=0^+$ or $0^-$ and $t>0$ one finds that $\mu=-2.0t$ and
this condition is satisfied. Therefore the bonding band is filled, and hence
the antibonding band is half filled, for an infinitesimal $\Delta$. Note that
this is not the case for $\Delta=0$ as we require at least an infinitesimal $\Delta$ to double
the size of the unit cell and cause the splitting into two bands, cf. Figs. \ref{fig:unit-cells}a and \ref{fig:unit-cells}b.
For $t<0$ the top of the bonding band is $\varepsilon_Y^-$ and we find the bonding band is filled for
 $\Delta\geq\Delta^<\equiv2\mu(\Delta^<)+4|t|$.
Solving this condition numerically we find that $\Delta^<=-0.64|t|$.

\begin{figure}
\epsfig{file=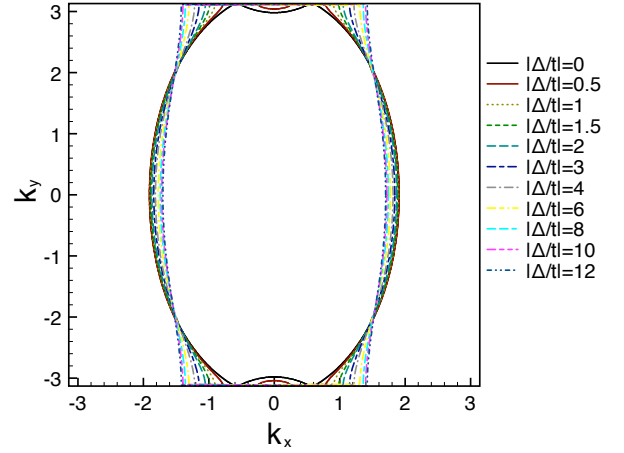,width=8cm} \caption{[Color
online] Fermi surface for non-interacting electrons for $t<0$. For small
\dt there are small Fermi pockets around the Y points. As \dt is increased the Fermi surface becomes increasingly quasi-one-dimensional.} \label{fig:ni-Fs-t<0}
\end{figure}

\begin{figure}
\epsfig{file=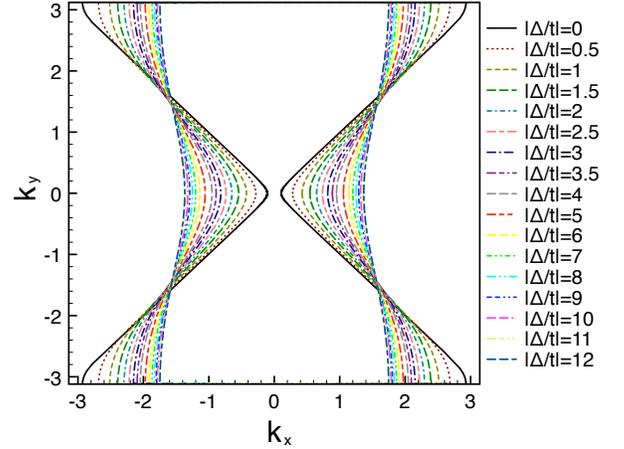,width=8cm} \caption{[Color
online] Fermi surface for non-interacting electrons  for $t>0$. For $t>0$ the bonding
band is always completely filled. Therefore this Fermi surface
arises entirely from the antibonding band. Note that the Fermi surface is very different from that for $t<0$ (Fig. \ref{fig:ni-Fs-t<0}).}
\label{fig:ni-Fs-t>0}
\end{figure}

In Figs. \ref{fig:ni-Fs-t<0} and \ref{fig:ni-Fs-t>0} we plot the Fermi surfaces for non-interacting electrons for each sign of $t$ at a range of \dtn. The first point to note is that at low \dt the two Fermi surfaces are very different. As \dt$\rightarrow\infty$ both Fermi surfaces become quasi-one-dimensional; but, even in  this limit the two Fermi surfaces are different in important ways as for $t<0$ the Fermi surface is electron like, while of $t>0$ the Fermi surface is hole like. Only when we reach the limit, and both Fermi surfaces are straight lines, are the two Fermi surfaces the same.

\begin{figure*}
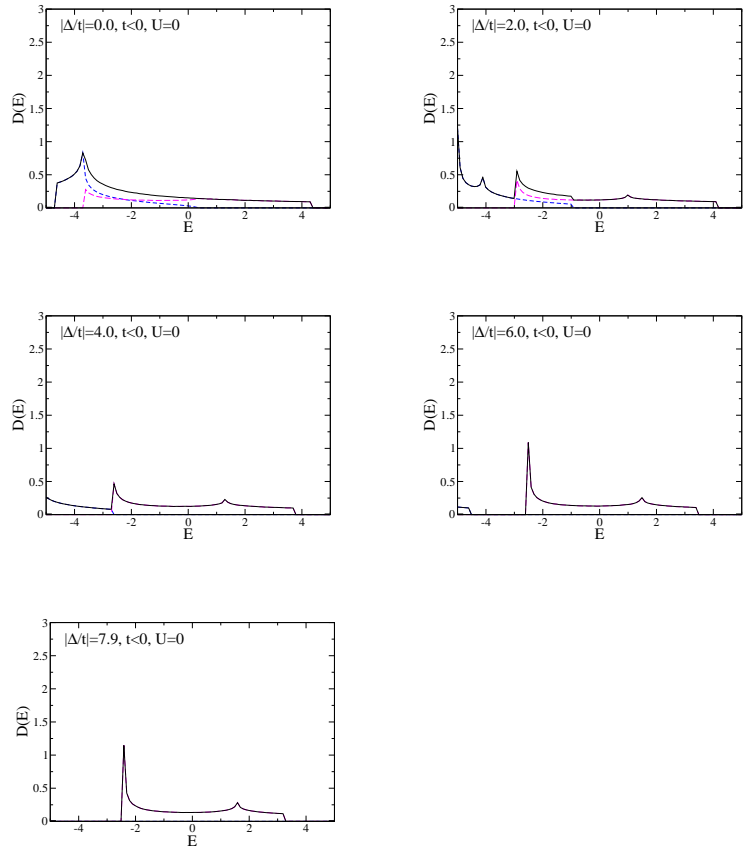

\epsfig{file=DOS_tlt0_Ueq0_Delta_eq0p000.eps,width=4.25cm} \hspace{1cm}
\epsfig{file=DOS_tlt0_Ueq0_Delta_eq2p000.eps,width=4.25cm}\vspace*{1cm}\\
\epsfig{file=DOS_tlt0_Ueq0_Delta_eq4p000.eps,width=4.25cm} \hspace{1cm}
\epsfig{file=DOS_tlt0_Ueq0_Delta_eq6p000.eps,width=4.25cm}\vspace*{1cm}\\
\epsfig{file=DOS_tlt0_Ueq0_Delta_eq7p900.eps,width=4.25cm} \hspace*{5.25cm}
\caption{[Color online] Density of states for $U=0$
and $t<0$. 
Dashed (blue and pink) lines indicate the
contributions of the individual bands.}\label{dos-lt}
\label{fig:DOS-t<0-U=0}
\end{figure*}

\begin{figure*}
\epsfig{file=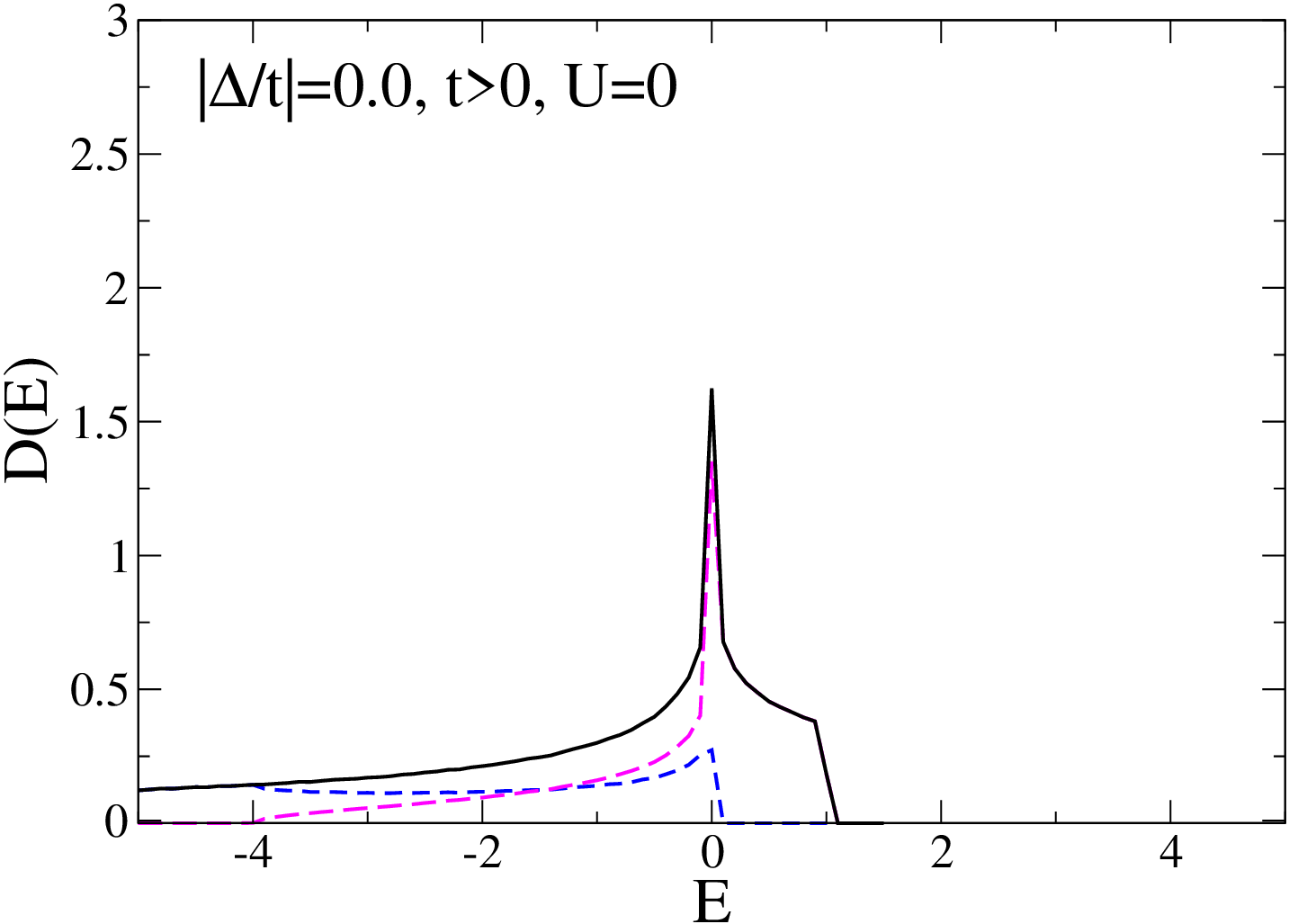,width=4.25cm} \hspace{1cm}
\epsfig{file=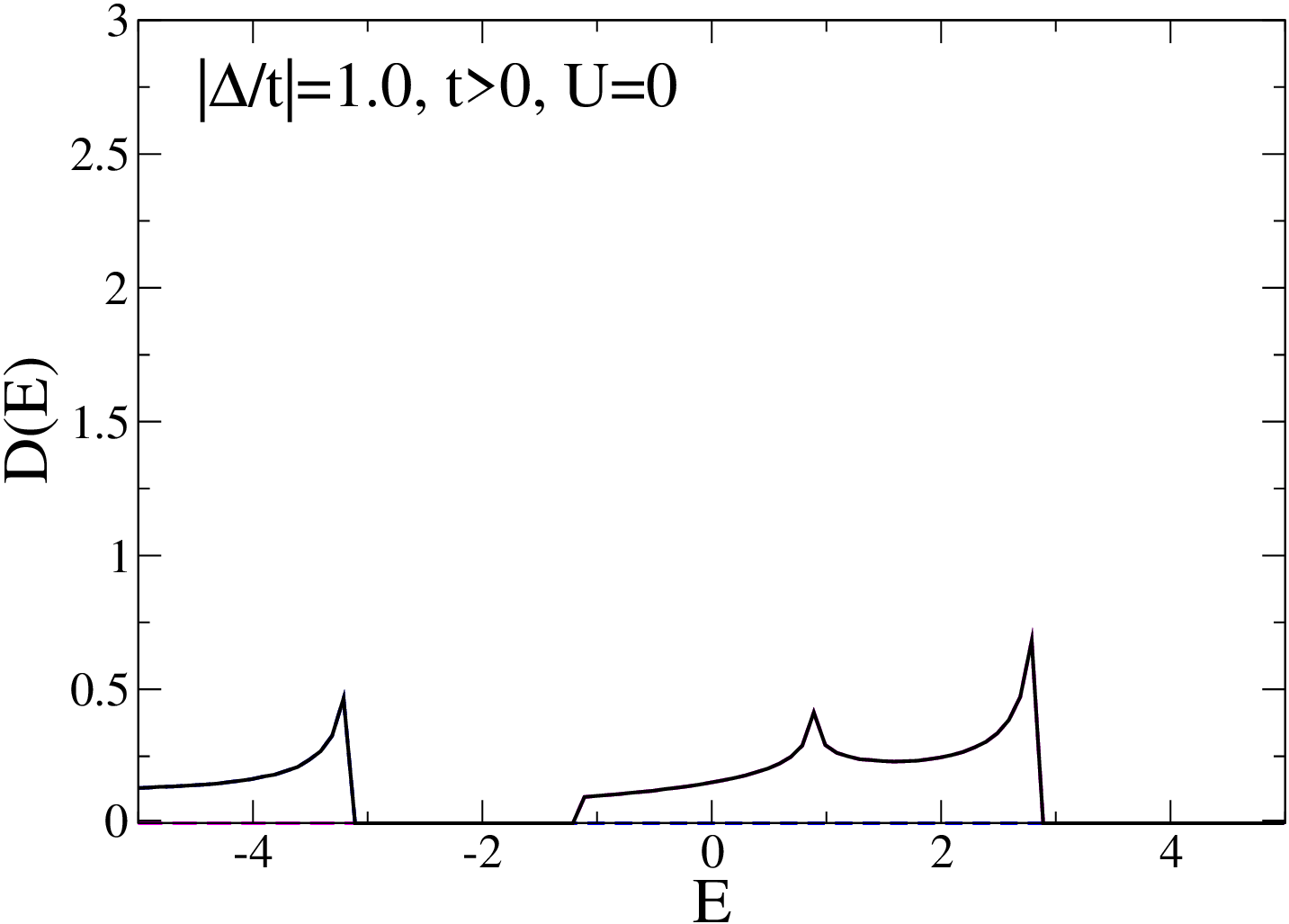,width=4.25cm}\vspace*{1cm}\\
\epsfig{file=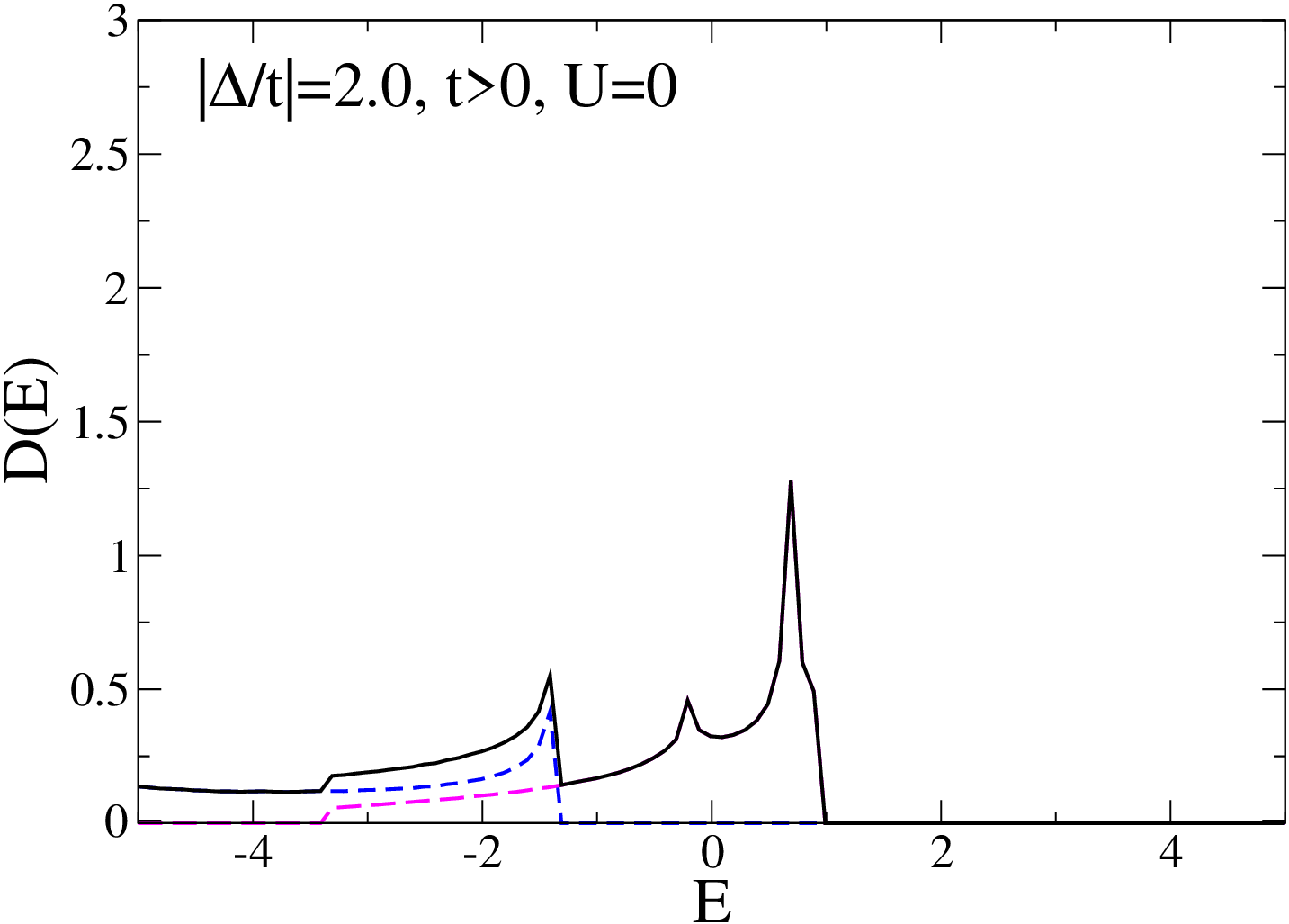,width=4.25cm} \hspace{1cm}
\epsfig{file=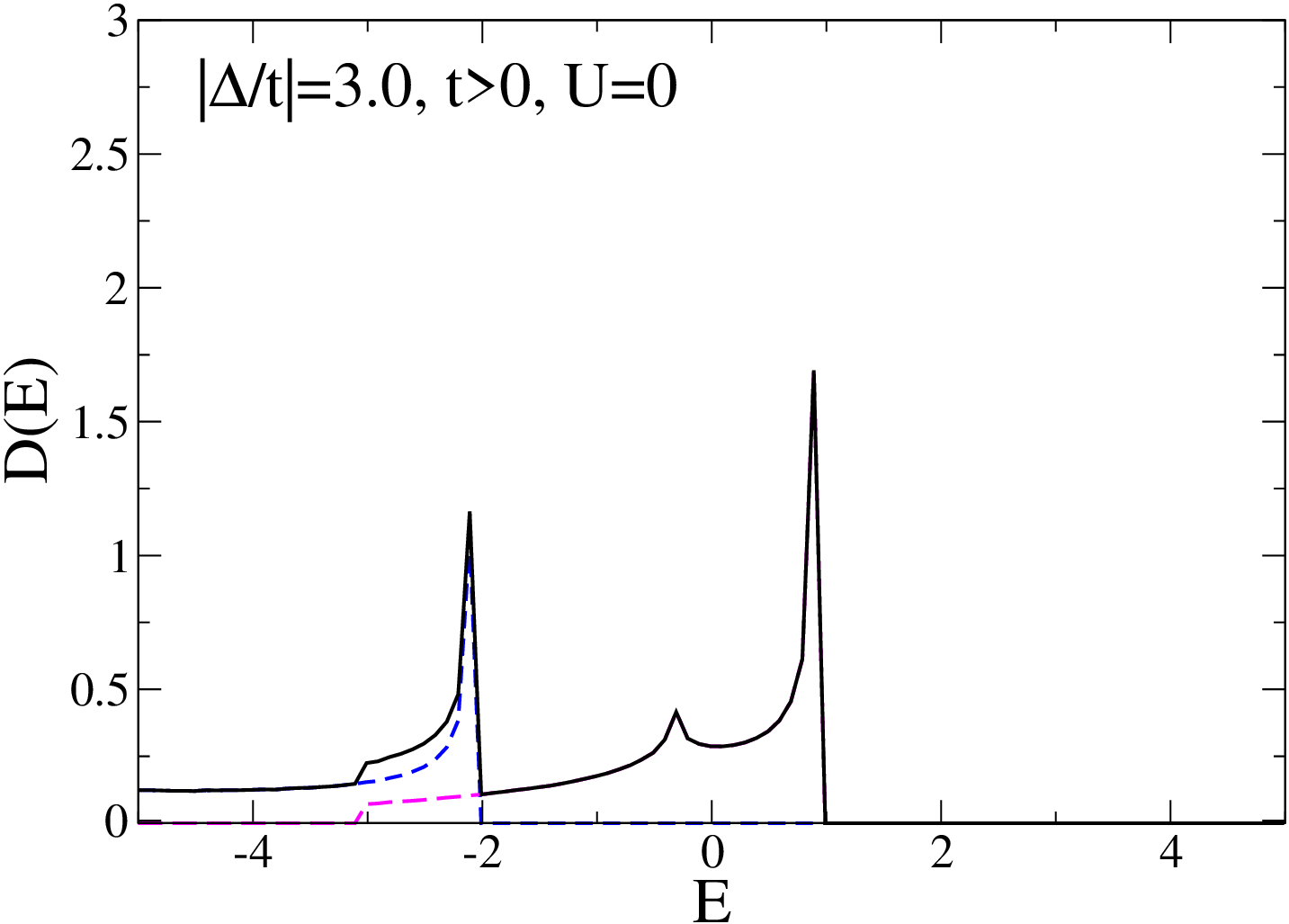,width=4.25cm}\vspace*{1cm}\\
\epsfig{file=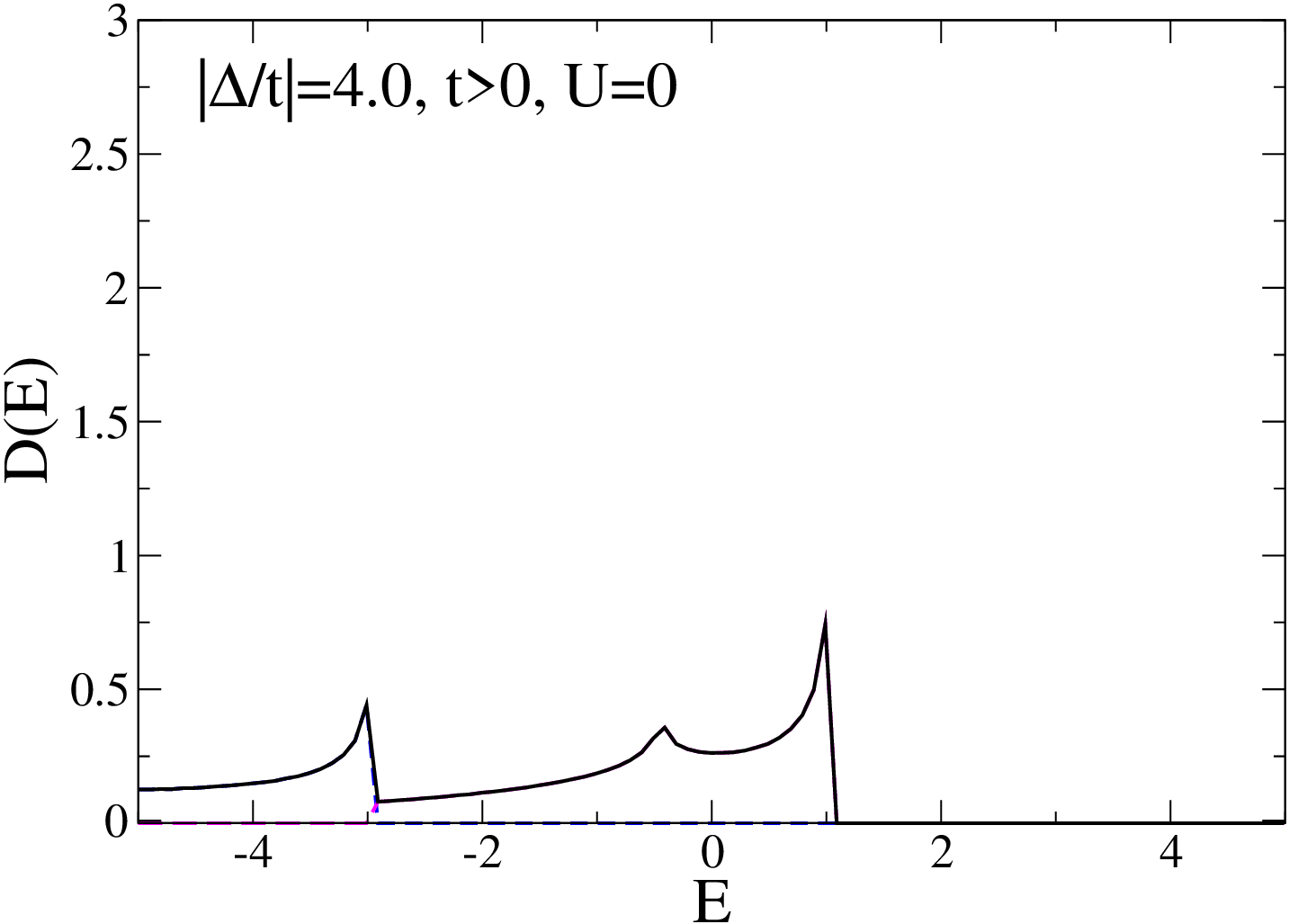,width=4.25cm} \hspace{1cm}
\epsfig{file=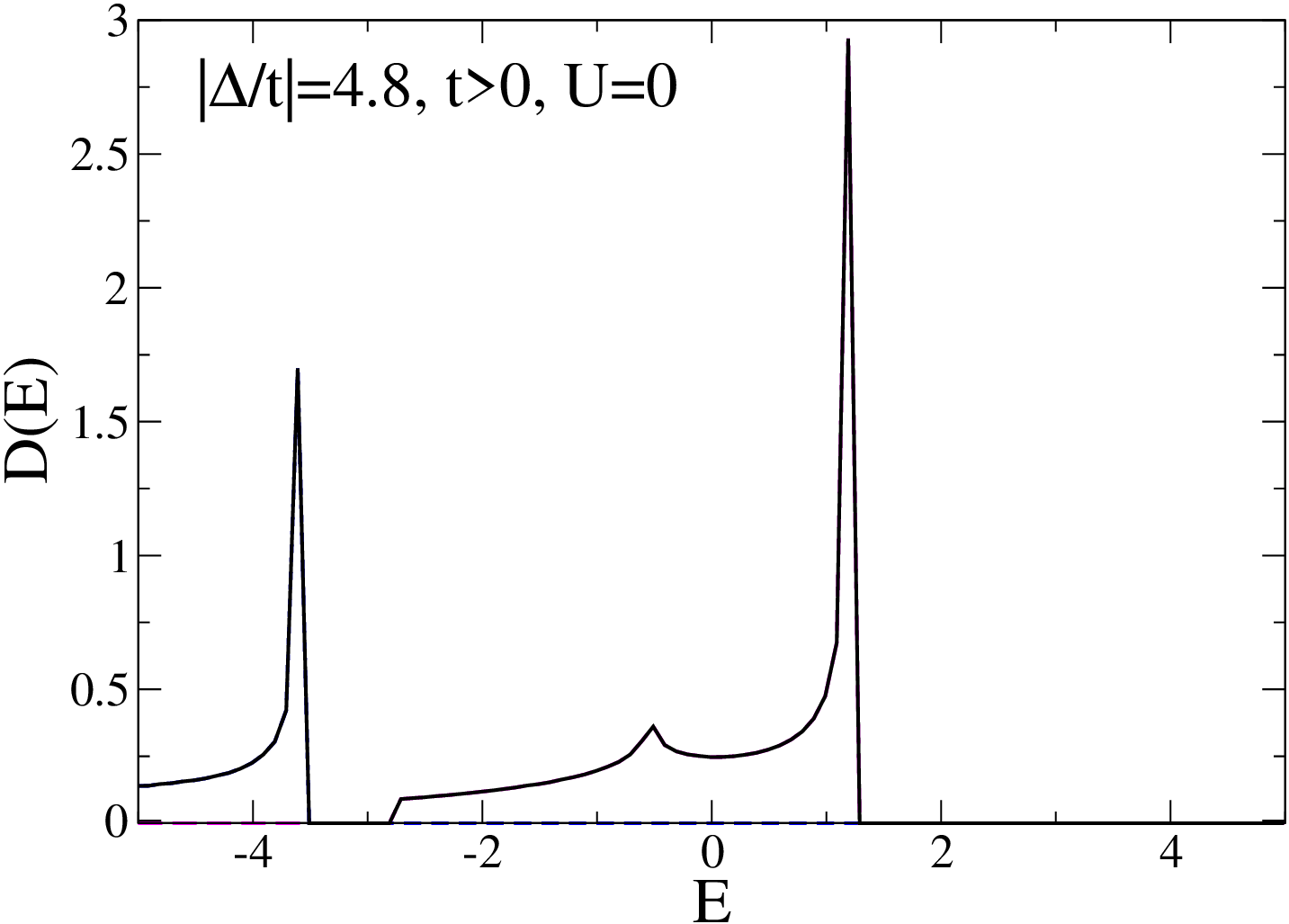,width=4.25cm}
\caption{[Color online] Density of states for $U=0$ and $t>0$. 
  Dashed (blue and
pink) lines indicate the contributions of the individual
bands.}\label{dos-gt} \label{fig:DOS-t>0-U=0}
\end{figure*}

We plot the DOS of the non-interacting ($U=0$) system in Figs.
\ref{dos-lt} and \ref{dos-gt} for $t<0$ and $t>0$ respectively.
Notice that the plots are very different for the two different signs
of $t$. In particular at $\Delta=0$ the Fermi energy is at the van
Hove singularity for $t>0$, but is not for $t<0$. This suggests that
the Fermi liquid will be less stable for $t>0$ than for $t<0$.
Indeed we find that, in the slave boson theory, the MIT occurs at
$|\Delta/t|=5.0$ for $t>0$ and $|\Delta/t|=8.0$ for $t<0$,
consistent with this expectation (cf. Fig. \ref{fig:q_-}). However, as \dt is increased the
van Hove singularity is moved away from the Fermi energy even for
$t>0$, so this,  weak coupling, description is clearly not the whole
story of the strong coupling theory.

\begin{thebibliography}{999}
\bibitem{aside}
 The dependence on the direction of the field may be
explained by (i) the larger g-factor for fields parallel to the
layers, and/or (ii) that the transitions at 51 K and 88 K have a
larger effect on the bulk magnetic susceptibility parallel to the
layers, $\chi_{ab}$ than on $\chi_{c}$ (see Fig. 1(c) in Ref.
\onlinecite{gasparovic} and Fig. 8 in  Ref. \onlinecite{watanabe}).
\end{thebibliography}
\end{document}